\documentclass[11pt]{article}
\usepackage{amsmath}
\usepackage{graphicx}
\usepackage{color}
\usepackage{orcidlink}
\usepackage{hyperref}
\hypersetup{colorlinks=true, linkcolor=blue, citecolor=blue, urlcolor=blue}
\usepackage{caption}
\usepackage{subcaption}
\RequirePackage[numbers,sort&compress]{natbib}
\begin{document}
\baselineskip=15pt

\begin{center}
    \LARGE{AdS Black String in a Cosmic Web: Geodesics, Shadows, and Thermodynamics}
\end{center}

\vspace{0.3cm}
\begin{center}
{\bf Faizuddin Ahmed\orcidlink{0000-0003-2196-9622}}\footnote{\bf faizuddinahmed15@gmail.com}\\
{\it Department of Physics, The Assam Royal Global University, Guwahati, 781035, Assam, India}\\
\vspace{0.15cm}
{\bf Ahmad Al-Badawi\orcidlink{0000-0002-3127-3453}}\footnote{\bf ahmadbadawi@ahu.edu.jo}\\
{\it Department of Physics, Al-Hussein Bin Talal University 71111, Ma’an, Jordan}\\
\vspace{0.15cm}
{\bf \.{I}zzet Sakall{\i}\orcidlink{0000-0001-7827-9476}}\footnote{\bf izzet.sakalli@emu.edu.tr (Corresp. author)}\\
{\it Physics Department, Eastern Mediterranean University, Famagusta 99628, North Cyprus via Mersin 10, Turkey}
\end{center}

\vspace{0.3cm}

\begin{abstract}
{\color{black}In a recent article (Ref. \cite{AOP}), the authors obtained a static, cylindrically symmetric Anti-de Sitter (AdS) black string (BS) solutions, which are cylindrical generalizations of black holes (BHs), surrounded by a cloud of strings (CS) and the quintessence field (QF), and discussed its properties. In the present study, we present a comprehensive analysis of cylindrically symmetric AdS BSs surrounded by CS and QF. Our analysis yields several significant results. We demonstrate that the presence of CS (parameter $\alpha$) and QF (parameters $c$ and $w$) reduces the radius of circular photon orbits (CPO) and BH shadow size, with measurements constrained by Event Horizon Telescope (EHT) observations of Sagittarius A*. We find that time-like particle orbits show increased energy with higher $\alpha$ and $c$ values. We calculate that C-energy, representing gravitational energy within cylindrical radius, decreases with radial distance but exhibits distinct responses to CS and QF parameters. We observe that scalar perturbation potential increases $r$ for specific $\alpha$ and $c$ values, indicating stronger field-spacetime interactions away from the BS. We determine that Hawking temperature increases linearly with Schwarzschild radius, with parameter-dependent behavior varying based on QF state parameter $w$. These results demonstrate how CS and QF significantly modify the geodesic, perturbative, and thermodynamic properties of AdS BS, with potential observational implications for gravitational lensing, accretion disk dynamics, and BH evaporation signatures in future astronomical observations. }

\end{abstract}


\section{Introduction}\label{sec:1}
 
Black holes (BHs) remain one of the most profound and intriguing predictions of general relativity (GR), and their study has led to transformative insights into the nature of spacetime, gravity, and quantum mechanics \cite{Bekenstein1,Hawking1,Wald1}. Among the various solutions in GR, a particularly interesting class is represented by cylindrically symmetric AdS BSs, which offer a valuable framework for investigating gravitational dynamics in both higher-dimensional and lower-dimensional contexts \cite{Lemos1,Lemos2,Cai1,Bronnikov1}. Unlike their spherically symmetric counterparts, BSs extend indefinitely along a spatial axis, which introduces significant differences in their structure and dynamics. This distinct feature makes BSs particularly relevant in the context of string theory, where they can provide a useful setting for studying the gravitational behavior of systems with extended spatial dimensions and for exploring the effects of external fields \cite{Mann1,Horowitz1,Chamblin1,Hendi1}.

The AdS spacetime associated with these solutions is especially important when considering the AdS/CFT (conformal field theory) correspondence, a powerful duality that connects gravitational theories in AdS space with CFTs on the boundary of AdS \cite{Maldacena1,Witten1,Gubser1,Aharony1}. This duality has profound implications for our understanding of quantum gravity, the thermodynamics of BHs, and strongly coupled systems, where the gravitational properties of BHs in AdS space inform various phenomena in quantum field theory.

A notable modification to the BH structure is the introduction of a CS surrounding the BH, which significantly impacts both the metric and thermodynamic properties of the BH system \cite{PSL1,Ghosh1,Toledo1,Bhatti1}. The concept of a CS arises from the classical Letelier model, which describes an ensemble of one-dimensional string-like objects that permeate spacetime and influence the gravitational field of the BH \cite{PSL2,PSL3,Rodrigues:2022zph,Singh:2020nwo,deMToledo:2018tjq}. These configurations offer a useful approximation for cosmic strings, which is hypothesized to be relics of phase transitions in the early universe, and may provide key insights into the formation of structure in the universe \cite{Vilenkin1, Vachaspati:1984g}. The presence of a CS alters the properties of the BH in a way that introduces significant deviations from the classical BH physics, including modifications to the event horizon, the geodesic motion, and the thermodynamic stability of the system. As a result, the CS parameter plays a critical role in shaping the geometry and dynamics of the BH, making it an essential tool for exploring alternative models of BH physics beyond the traditional framework \cite{Eslam1,Liang:2020uul,Rani:2024qju}.

Another crucial factor influencing the gravitational field of BHs is QF, a theoretical form of dark energy that is characterized by a dynamical equation of state, with the state parameter $w$ ranging from $w = \frac{1}{3}$ to $w = -1$ \cite{Kiselev1,Bronnikov2,Mota1,Singh1}. Unlike the cosmological constant, which is static, QF allows for a time-varying energy density and plays a significant role in modifying the spacetime curvature. This modification can lead to distinctive thermodynamic behavior, including alterations to the BH's entropy, temperature, and stability. Furthermore, the presence of quintessence can influence gravitational phenomena such as lensing, stability of orbits, and geodesic structures in the vicinity of the BH \cite{MoraisGraca:2016hmv,Gogoi:2024ypn,Tsilioukas:2024seh}. The study of BHs in the presence of quintessence is essential for deepening our understanding of the interplay between dark energy and the strong gravitational fields of BHs, as well as for testing various alternative cosmological models \cite{Hamil:2023dmx,Tan:2024sgv,Li:2023ecv,Malligawad:2024uwi}. {\color{black}When both CS and QF are present around a BS, their combined effects introduce complex interactions that significantly modify the spacetime geometry beyond what either effect would produce independently \cite{Fernando1,Zhang1,Ovgun1,Parbin1}. The CS, characterized by the parameter $\alpha$ with $0 < \alpha < 1$, acts predominantly to alter the local geometry by introducing a deficit angle, effectively reducing the circumference-to-radius ratio around the symmetry axis. Meanwhile, the QF, parameterized by the state parameter $w$ (ranging from $w = \frac{1}{3}$ to $w = -1$) and the normalization constant $c$, primarily influences the large-scale structure of spacetime through its negative pressure. When these effects combine, they produce a rich interplay that enhances certain features while suppressing others. For instance, our analysis demonstrates that while both parameters individually reduce the COP radius, their combined effect produces a more pronounced reduction than the sum of their individual contributions. Similarly, the Hawking temperature's response to these parameters exhibits a non-linear relationship where the effect of $w$ can either amplify or diminish the impact of $\alpha$, depending on its specific value. This non-trivial interaction between CS and QF parameters offers a uniquely valuable framework for exploring modified gravitational effects in AdS backgrounds, where boundary conditions are critical in determining system properties. Understanding these combined effects is essential for mapping the full parameter space of possible BS solutions with external matter fields, which may have significant implications for theoretical models in quantum gravity and cosmology.}

The main motivation behind this work is to build upon recent studies of static, cylindrically symmetric AdS BS surrounded by both a cloud of strings and quintessence \cite{AOP}. Although earlier work has focused primarily on deriving these solutions and investigating their fundamental thermodynamic properties, a thorough examination of the geodesic motion, scalar perturbations, and detailed thermodynamics is still lacking. For example, studies of geodesic motion around a compact object are crucial for understanding the stability of circular orbits and the dynamics of test particles (see, for example, Refs. \cite{NPB, CJPHY, AHEP1, AHEP2, EPJC}). Similarly, detailed perturbation analyses, particularly scalar perturbations (see, for example, Refs. \cite{NPB,CJPHY,AHEP1,AHEP2,EPJC} and related references there in), could provide insights into the stability of these solutions and their response to external perturbations. {\color{black} Our comprehensive approach addresses critical gaps by examining geodesic motion, scalar perturbations, and detailed thermodynamics together. These analyses are complementary and necessary for a complete understanding of BS physics with CS and QF. Geodesic studies reveal spacetime's causal structure and observable phenomena like BH shadows, while our analysis of circular orbits addresses fundamental questions about accretion and system stability. Scalar perturbations provide key insights into wave propagation and field-geometry interactions. Combined with our thermodynamic analysis connecting quantum properties to macroscopic observables, this approach characterizes how CS and QF modify BS behavior across all relevant physical aspects, creating a coherent picture of how external matter fields alter BS physics. Therefore,} this work aims to fill these gaps by investigating the following key aspects:

1. We investigate the shadow and geodesic motion of both massless and massive test particles near the AdS BS solution. By deriving the effective potential, we identify the radius of CPO and calculate the BH shadow radius. Additionally, we analyze the forces acting on photons within this background. Moreover, we examine the motion of time-like particles, focusing on their circular orbits, and study how the CS and QF parameters influence key quantities such as energy, angular momentum, and angular velocity \cite{Mustafa:2021,Li:2023,Delsate:2015,Ramos:2021,Bardeen:1972,sharma1,abdu1,chen1,kala1}.

2. We analyze the perturbative potential for a massless scalar field propagating in the BS background. Using the Klein-Gordon equation, we investigate how the presence of CS and QF modifies the perturbative structure, influencing the stability and wave propagation within this geometry \cite{Konoplya:2011,Cardoso:2009,Regge:1957,Kokkotas:1999,Molina:2016,Zinhailo:2019}.

3. We explore the thermodynamic properties of the BS solution, focusing on how the CS and QF parameters influence key thermodynamic quantities, including the Hawking temperature, entropy, and C-energy. These analyses offer valuable insights into the interplay between quantum gravity effects and the thermodynamic behavior of the BS in an AdS background \cite{Hawking:1974rv,Bekenstein:1973ur,Wald:1993nt,Page:2005xp,sakalli:2022wy,Cai:1996eg,Chamblin:1999tk}.

The paper is organized as follows. In Sec. \ref{sec:2}, we introduce the AdS BS solution surrounded by both a CS and a QF, discussing its fundamental properties. In sec. \ref{sec:3}, we conduct a geodesic analysis of both light-like and time-like particles, deriving the effective potentials and investigating the stability conditions. Sec. \ref{sec:4} focuses on the analysis of the C-energy associated with the BS solution, which serves as a measure of the gravitational energy contained within a given cylindrical radius. Section \ref{sec:5} examines the perturbative structure of a massless scalar field in this background, exploring its implications for the stability of the system. In Sec. \ref{sec:6}, we investigate the thermodynamic properties of the BS, including the Hawking temperature, entropy, and their physical implications. Finally, Sec. \ref{sec:7} concludes the paper, summarizing our findings and suggesting potential directions for future research.

\section{Cylindrically symmetric A\lowercase{d}S BS surrounded by CS and QF: Features and geodesics} \label{sec:2}

In a recent article, Ref. \cite{AOP}, the authors derived a static, cylindrically symmetric AdS BS solution surrounded by CS and QF, and discussed its properties. In the present study, we extend this work by examining the behavior of light-like and time-like particles in the vicinity of this AdS BS spacetime, with a detailed analysis of their geodesic motion. Moreover, we determine $C$-energy of this AdS BS solution and show how CS as well as QF influence the C-energy for a specific state parameter. In addition, we investigate scalar perturbations through the massless Klein-Gordon equation and discuss the outcomes showing the effects of CS and QF. Finally, we study thermal properties associated with this BS solution and analyze the results.

The metric of a static and cylindrically symmetric AdS BS surrounded by CS and QF is described by the following line-element \cite{AOP} {\color{black} (in the natural units, where $c=1=\hbar=G$)}
\begin{eqnarray}
&&ds^2=-A(r)\,dt^2+\frac{dr^2}{A(r)}+r^2\,d\varphi^2+\frac{r^2}{\ell^2_{p}}\,dz^2,\label{aa1}\\
&& -\infty < t < +\infty,\quad\quad r\geq 0,\quad\quad \varphi \in [0, 2\,\pi),\quad\quad -\infty < z < +\infty,\nonumber
\end{eqnarray}
with the metric function $A(r)$ given by
\begin{eqnarray}
A(r)=\alpha-\frac{2\,M}{r}+\frac{r^2}{\ell^2_{p}}+\frac{c}{r^{3\,w+1}},\quad\quad -1 < w < -\frac{1}{3},\label{aa2}
\end{eqnarray}
where $0 < \alpha <1$ is the CS parameter \cite{PSL1}, $M$ denotes mass of BS, $w$ represents the state parameter of the quintessence matter and $c\geq 0$ serves as a positive constant, and $\ell_p$ represents AdS radius, related with the cosmological constant $\frac{1}{\ell^2_{p}}=-\frac{\Lambda}{3}$.

Now, taking ordinary derivative of $A(r)$ with respect to $r$ and then multiplying both side by $r$ results
\begin{eqnarray}
    r\,A'(r)=\frac{2\,M}{r}+\frac{2\,r^2}{\ell^2_{p}}-\frac{c\,(3\,w+1)}{r^{3\,w+1}}.\label{aa3}
\end{eqnarray}
Therefore, we find the quantity $(2\,A-r\,A')$ as follows:
\begin{equation}
    2\,A(r)-r\,A'(r)=2\,\left(\alpha-\frac{3\,M}{r}+\frac{c\,(3\,w+3)}{2\,r^{3\,w+1}}  \right).\label{aa4}
\end{equation}

\subsection{\large \bf Geodesics Motions of AdS BS Spacetime}\label{sec:3}


In this study, we explore the motion of photons and massive particles in the gravitational field of the anti-de Sitter black string solution described by Eq. (\ref{bb1}). Our primary objective is to analyze how different parameters, particularly the cosmic string ($\alpha$) and the QF parameters $(c, w)$, influence the dynamics of test particles. Geodesic analysis has proven to be a powerful method in general relativity for probing the structure of spacetime and the physical properties of compact objects. In spherically symmetric black hole spacetimes, both singular and regular, geodesic motion has been widely studied, yielding insights into particle trajectories and gravitational lensing (see Refs. \cite{NPB,CJPHY,AHEP1,AHEP2,EPJC} and references therein).

{\color{black}
However, geodesic motion in cylindrically symmetric spacetimes introduces distinct features not present in spherical geometries. Cylindrical symmetry implies invariance under rotations about and translations along a central axis, typically the $z$-axis. This symmetry is relevant in a variety of physical contexts, such as cosmic strings, rotating cylindrical configurations, and relativistic fluid flows. Analyzing the geodesics in such backgrounds provides insight into the influence of anisotropic geometry and topological features on particle dynamics. Moreover, it offers potential observational signatures associated with non-spherical compact objects or field configurations. For further discussion on geodesic motion in cylindrical spacetimes, the reader is referred to Refs. \cite{LH1,LH2,LH3,LH4}.
}

To study non(-planar) geodesics motion, we begin by writing the Lagrangian density function using the metric (\ref{aa1}) given by
\begin{equation}
    \mathcal{L}=\frac{1}{2}\left[-A(r)\,\dot{t}^2+\frac{\dot{r}^2}{A(r)}+r^2\,\dot{\varphi}^2+\frac{r^2}{\ell^2_{p}}\,\dot{z}^2\right],\label{bb1}
\end{equation}
where the dot represents ordinary derivative w. r. t. an affine parameter. Planar geodesic means $\dot{z}=0$ and $\dot{z} \neq 0$ for non-planar.

From the Lagrangian density function given in Eq. (\ref{bb1}), we observe that this function depends only on the axial distance $\mathcal{L}=\mathcal{L}(r)$ and is independent of the temporal ($t$), angular ($\varphi$), and translational ($z$) coordinates. As a result, there are three conserved quantities associated with these cyclic coordinates. These are, namely, the energy $\mathrm{E}$ given by
\begin{equation}
    \mathrm{E}=-\frac{\partial \mathcal{L}}{\partial \dot{t}}\Rightarrow \mathrm{E}=A(r)\,\dot{t}\Rightarrow \dot{t}=\frac{\mathrm{E}}{A(r)}.\label{bb2}
\end{equation}
The angular momentum $\mathrm{L}$ about $z$-axis is given by
\begin{equation}
    \mathrm{L}=\frac{\partial \mathcal{L}}{\partial \dot{\varphi}}\Rightarrow \mathrm{L}=r^2\,\dot{\varphi}\Rightarrow \dot{\varphi}=\frac{\mathrm{L}}{r^2}.\label{bb3}
\end{equation}
And the linear momentum along the $z$-axis is given by
\begin{equation}
    p_z=\frac{\partial \mathcal{L}}{\partial \dot{z}}\Rightarrow p_z=\frac{r^2}{\ell^2_{p}}\,\dot{z}\Rightarrow \dot{z}=\frac{\ell^2_{p}}{r^2}\,p_z.\label{bb3aa}
\end{equation}

With these, the geodesic equation for the axial coordinate $r$ using Eq. (\ref{bb1}) can be rewritten as
\begin{equation}
    \dot{r}^2+V_\text{eff}(r)=\mathrm{E}^2\label{bb4}
\end{equation}
which is equivalent to the one-dimensional equation of motion of test particles of unit mass having energy $\mathrm{E}^2$ and the effective potential $V_\text{eff}$ given by
\begin{eqnarray}
    V_\text{eff}&=&\left(-\varepsilon+\frac{\mathrm{L}^2}{r^2}+\frac{\ell^2_{p}}{r^2}\,p^2_{z}\right)\,A(r)=\left(-\varepsilon+\frac{\mathrm{L}^2}{r^2}+\frac{\ell^2_{p}}{r^2}\,p^2_{z}\right)\,\left(\alpha-\frac{2\,M}{r}+\frac{r^2}{\ell^2_{p}}+\frac{c}{r^{3\,w+1}}\right),\label{bb5}
\end{eqnarray}
where $\varepsilon=2\,\mathcal{L}=0$ for null geodesics and $-1$ for time-like particles.

Equation (\ref{bb5}) shows that the effective potential for both light-like and time-like particles, is controlled by several factors involved in the BH spacetime. Among these are the CS parameter $\alpha$, the positive normalization constant $c$ associated with QF, state parameter $w$ of QF, and the AdS radius $\ell_p$. Additionally, the effective potential is further modified by the momenta $(\mathrm{L}, p_z)$ and the BH mass $M$. These parameters collectively shape the underlying gravitational field of the chosen BS solution. As a result, they induce alterations in the dynamics of light-like and time-like particles, influencing their motion in a non-trivial manner. The interplay between these factors leads to a rich behavior in the spacetime geometry, affecting the trajectory and overall motion of test particles in such a spacetime environment. 

\begin{center}
    {\bf Special case: Planar geodesics, $\dot{z}=0$}
\end{center}

For planar geodesics, we have $\dot{z}=0$ which implies $p_z=0$. In that case, the effective potential for null and time-like geodesics reduces as
\begin{eqnarray}
    V_\text{eff}&=&\left(-\varepsilon+\frac{\mathrm{L}^2}{r^2}\right)\,A(r)=\left(-\varepsilon+\frac{\mathrm{L}^2}{r^2}\right)\,\left(\alpha-\frac{2\,M}{r}+\frac{r^2}{\ell^2_{p}}+\frac{c}{r^{3\,w+1}}\right),\label{nn1}
\end{eqnarray}

Below, we discuss in detail the motions of light-like and time-like particles and analyze the outcomes.

\subsubsection{\bf Null geodesics}

For non-planar null geodesics, we have $\varepsilon=0$. The effective potential thus from Eq. (\ref{bb5}) becomes
\begin{equation}
    V_\text{eff}=\frac{\lambda^2}{r^2}\,\left(\alpha-\frac{2\,M}{r}+\frac{r^2}{\ell^2_{p}}+\frac{c}{r^{3\,w+1}}\right),\quad\quad \lambda=\sqrt{\mathrm{L}^2+\ell^2_{p}\,p^2_{z}}.\label{cc1}
\end{equation}

\begin{figure}[ht!]
    \centering
    \subfloat[$\alpha=0.1$]{\centering{}\includegraphics[width=0.48\linewidth]{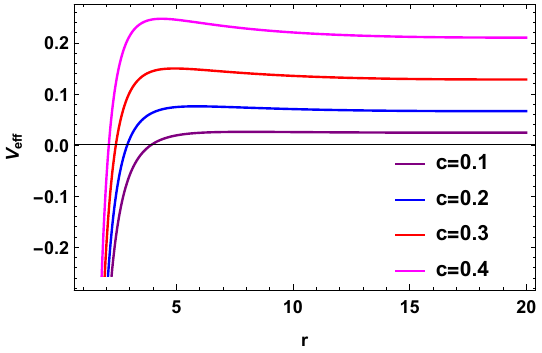}}\quad
    \subfloat[$c=0.2$]{\centering{}\includegraphics[width=0.48\linewidth]{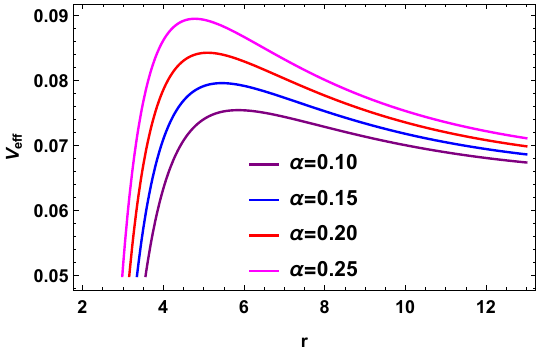}}\\
    \subfloat[]{\centering{}\includegraphics[width=0.48\linewidth]{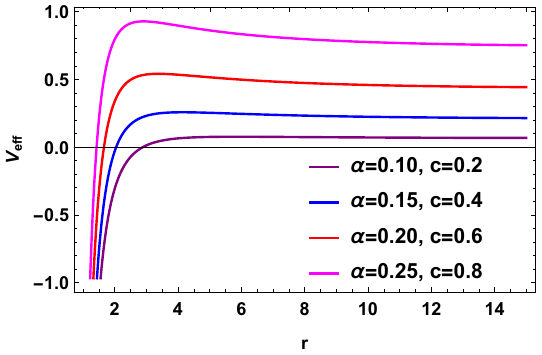}}
    \caption{The effective potential for non-planar null geodesics, $\dot{z} \neq 0$. Here, $\Lambda = -0.003$, $\mathrm{L} = 1$, $\mathrm{p}_z = 0.01$, and $M = 1$.}
    \label{fig:null}
\end{figure}

{\color{black} It is well-known that the effective potential for null geodesics around a BH is a tool used to describe the behavior of light (or massless particles) in the curved spacetime near a BH. It helps us understand the motion of light in regions with highly curved spacetime, such as near the event horizon. The effective potential provides insight into whether photons can escape the BH's gravitational pull (if they are outside the CPO) or whether they will be captured by the BH.} In Figure \ref{fig:null}, we present the effective potential for non-planar null geodesics given in Eq. (\ref{cc1}) with varying values of the CS parameter $\alpha$ and the quintessence constant $c$, for a specific state parameter $w = -2/3$, the cosmological constant $\Lambda = -0.003$, the angular momentum $\mathrm{L} = 1$, the translational momentum $\mathrm{p}_z = 0.01$, and the BH mass $M = 1$. It is observed that as {\color{black} the CS parameter $\alpha$, the QF parameter $c$ and their combination ($\alpha,c$) increase, the effective potential for photon light in the gravitational field produced by the BH also increases with increasing $r$. This indicates that the influence of the gravitational field on the motion of photon light increases with larger values of these parameters and may be captured by the BH.}

Now, we study circular null geodesics of radius $r=r_c$ in the constant $z$-plane ($p_z=0$) in the same sense that we define the equatorial plane in spherical symmetry. Thus, we have the conditions $\dot{r}=0$ and $\ddot{r}=0$ which implies the following relations using Eq. (\ref{bb4}) as
\begin{equation}
    V_\text{eff}(r)=\mathrm{E}^2,\quad\quad V'_\text{eff}(r)=0,\label{special1}
\end{equation}
where prime denotes partial derivative w. r. t. $r$.

The first relation $V_\text{eff}(r)=\mathrm{E}^2$ can be rewritten using  Eq. (\ref{cc1}) as follows:
\begin{equation}
    \mathrm{E}^2=\frac{\mathrm{L}^2}{r^2}\,\left(\alpha-\frac{2\,M}{r}+\frac{r^2}{\ell^2_{p}}+\frac{c}{r^{3\,w+1}}\right).\label{special2}
\end{equation}
Hence, the critical impact parameter ($\beta_c=\mathrm{L}/\mathrm{E}$) for photon particles in the constant $z$-plane can be determined as follows:
\begin{equation}
    \frac{1}{\beta^2_c}=\frac{\mathrm{E}^2}{\mathrm{L}^2}=\frac{1}{r^2_c}\left(\alpha-\frac{2\,M}{r_c}+\frac{r^2_c}{\ell^2_{p}}+\frac{c}{r^{3\,w+1}_c}\right).\label{special3}
\end{equation}

The expression given in Eq. (\ref{special3}) shows that the critical impact parameter for photon particles is influenced by several factors. These include the CS parameter $\alpha$, the QF parameters $(c, w)$, the AdS radius $\ell_p$, and  the BH mass $M$, which all together alter the impact parameter and consequently the motion of photons in the BS spacetime.

Now, with the help of the effective potential for null geodesics given in Eq. (\ref{cc1}) in the constant $z$-plane (planar null geodesics, $\dot{z}=0$), we explain phenomena like the radius of circular photon orbits (CPO)\footnote[4]{In cylindrical symmetry, the concept of a CPO differs from that in spherically symmetric spacetimes due to the unique nature of the gravitational field, which is symmetric around the $z$-axis rather than being isotropic \cite{bsr1}. This means the gravitational field remains uniform in the azimuthal $\varphi$-direction and varies with the radial distance $r$. Consequently, there is no single, well-defined photon sphere as in spherical symmetry. However, a notable feature in such spacetimes is the possibility of CPO, where photons may follow circular paths around the $z$-axis. The existence and properties of such orbits are highly dependent on the specific characteristics of the gravitational field-for example, whether the source is rotating, charged, or has other properties consistent with cylindrical symmetry, such as in the case of a rotating cylinder or a black hole with cylindrical geometry.} and BH shadows and demonstrate how various parameters involved in the spacetime geometry influences these. This can be determined by the relation $V'_\text{eff}(r=r_\text{ph})=0$. Using expression (\ref{cc1}) in the constant $z$-plane, we find the following relation:
\begin{equation}
    r\,A'(r)-2\,A(r)=0.\label{cc2}
\end{equation}
Substituting Eq. (\ref{aa4}) into the Eq. (\ref{cc2}), we find the following relations:
\begin{equation}
    \alpha-\frac{3\,M}{r}+\frac{3\,c\,(w+1)/2}{r^{3\,w+1}} =0.\label{cc3}
\end{equation}
The above equation is complicated to solve without choosing state parameter $w$. We can see that radius of CPO is alter by the CS parameter $\alpha$, and the quintessence matter field parameters $(c, w)$. 

To determine radius of CPO for a given state parameter, we assign $w=-2/3$. Consequently, Eq. (\ref{cc3}) with $w=-2/3$ becomes
\begin{equation}
    \alpha-\frac{3\,M}{r}+\frac{c\,r}{2} =0\Rightarrow r_\text{ph}=\frac{\alpha}{c}\,\left[-1+\sqrt{1+\frac{6\,c\,M}{\alpha^2}}\right].\label{cc4}
\end{equation}
In the limit where $\alpha=0$, that is, no CS parameter, for the state parameter, $w=-2/3$, the radius of CPO from Eq. (\ref{cc3}) becomes
\begin{equation}
    r_\text{ph}=\sqrt{\frac{6\,M}{c}}.\label{cc5aa}
\end{equation}

\begin{figure}[ht!]
    \centering
    \subfloat[$c=0.02$]{\centering{}\includegraphics[width=0.48\linewidth]{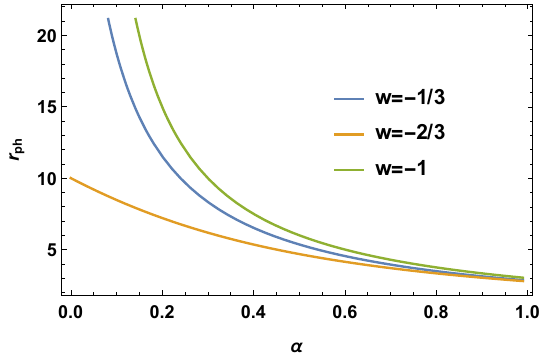}}\quad
    \subfloat[$w=-2/3$]{\centering{}\includegraphics[width=0.48\linewidth]{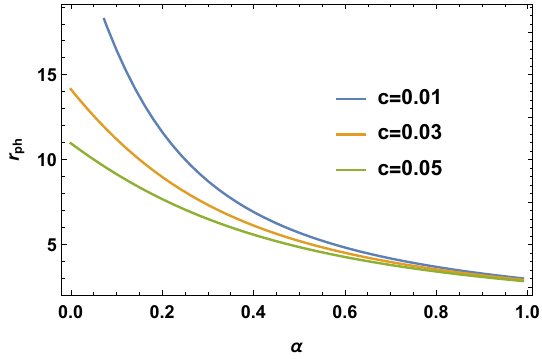}}
    \caption{ Variation of the radius of CPO with $\alpha$, for different values of $w$ (left panel) and $c$ (right panel). Here with set $M=1$.}
    \label{figa1}
\end{figure}

Another state parameter, for example $w=-1/3$, we find from Eq. (\ref{cc3}) the radius of CPO as,
\begin{equation}
    r_\text{ph}=\frac{3\,M}{\alpha+c}.\label{cc5bb}
\end{equation}
In the limit where $\alpha=0$, that is, no CS parameter, we find the radius
\begin{equation}
    r_\text{ph}=\frac{3\,M}{c}.\label{cc5cc}
\end{equation}
Moreover, in the limit where $c=0$, with no QF, we find the radius
\begin{equation}
    r_\text{ph}=\frac{3\,M}{\alpha},\label{cc5dd}
\end{equation}
which is the same radius for the state parameter $w=-1$ using Eq. (\ref{cc3}). The radius of CPO is now represented by a figure showing the influence of CS and QF parameters. Figure \ref{figa1} illustrates that radius of CPO shrinks with both $\alpha$ and $c$ values. However, $\alpha$ exerts a bigger influence on the radius of CPO than $c$.

\begin{figure}[ht!]
    \centering
    \subfloat[$c=0.02$]{\centering{}\includegraphics[width=0.48\linewidth]{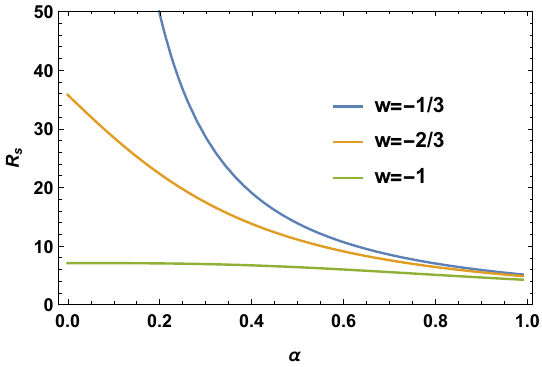}}\quad
    \subfloat[$w=-2/3$]{\centering{}\includegraphics[width=0.48\linewidth]{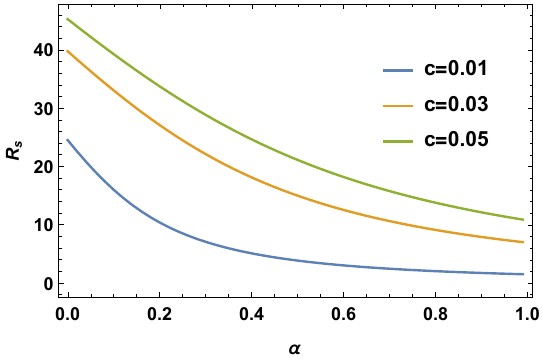}}
    \caption{ Variation of the shadow radius with $\alpha$, for different values of $w$ (left panel) and $c$ (right panel). Here we set $M=1$ and $\ell_p=300$.}
    \label{figa2}
\end{figure}

The BS shadow radius $R_s$, as perceived by an observer at infinity, is defined by \begin{equation} R_s = \frac{r_\text{ph}}{\sqrt{A(r_\text{ph})}} .\label{shadeq1}
\end{equation}
We provide analytical solutions for the shadow radius of the  BS  based on the metric (\ref{aa1}) 
 \begin{eqnarray}
  &&(i)\,\,\mbox{for}\,\, w=-1/3, \,\,R_{s}= \frac{3\,\sqrt{3}\,M}{\sqrt{(\alpha+c)^2+\frac{27\,M^2}{\ell^2_p}}},     
 \\
&&(ii)\,\,\mbox{for}\, \, w=-2/3, \,\,R_{s}= \frac{\alpha\,\left(-1+\sqrt{1+\frac{6\,c\,M}{\alpha^2}}\right)}{c\sqrt{\frac{\alpha}{3}\left(-1+2\sqrt{1+\frac{6\,c\,M}{\alpha^2}}\right)+\frac{6\,c\,M-2\, \alpha^2\left(-1+\sqrt{1+\frac{6\,c\,M}{\alpha^2}}\right)}{c^2\, \ell^2_p}}},\\
&&(iii)\,\,\mbox{for}\,\, w=-1,\,\,R_{s}=   \frac{3\,M}{\alpha\,\sqrt{\frac{9\,(1+c\,\ell^2_p)\,M^2}{\alpha^2\,\ell^2_p}+\frac{\alpha}{3}}}.     \label{bb16cc}
\end{eqnarray}
The preceding equations describe the influence of BS parameters on photon trajectories within the BS spacetime. To better understand this effect, we present numerical values for the shadow radius. Table \ref{table2a} illustrates how the parameter $\alpha$ and $c$ reduce the shadow radius. Furthermore, as $w$ increases, the shadow radius decreases. Figure \ref{figa2} displays the relationship between the shadow radius and the parameter $\alpha$, where the $R_{s}$ values align with those in Table \ref{table2a}. {\color{black} The EHT collaboration reported a shadow angular diameter of $\theta_{\text{sh}} = 48.7 \pm 7.0 \mu$as for Sgr A* \cite{EventHorizonTelescope:2022xqj}. To apply constraints to our model, we follow Vagnozzi et al.'s \cite{Vagnozzi:2022moj} approach by converting our theoretical shadow radius $R_s$ to an observable angular diameter using $\theta_{\text{sh}} = 2R_s(M/D)$, where $D$ is the distance to Sgr A* (approximately $8.15 \pm 0.15$ kpc). Comparing our calculated shadow radii in Table \ref{table2a} with this observational constraint, and accounting for the mass-to-distance ratio of Sgr A* ($M/D \approx 4.3 \times 10^{-11}$ radians), we found that for $w=-2/3$ and $c=0.001$, values of $\alpha$ near 0.6 yielding $R_s \approx 9.97$ would produce angular diameters most compatible with EHT observations. Similarly, for $w=-1$ with the same $c$ value, $\alpha$ values around 0.4-0.6 (with $R_s$ ranging from 7.45 to 8.98) align well with the observed constraints. Other parameter combinations produce either too large or too small shadow sizes compared to observations. This analysis represents an important connection between our theoretical model and observational data, demonstrating how current and future EHT observations can place meaningful constraints on modified gravity theories involving CS and QF \cite{EventHorizonTelescope:2022xqj, Vagnozzi:2022moj,Junior:2024vdk}.}

\begin{center}
\begin{tabular}{|c|c|c|c|c|c|c|c|c|c|}
 \hline  \multicolumn{10}{|c|}{ $R_s$} \\ \hline
  & \multicolumn{3}{|c|}{ $w=-1/3$}& \multicolumn{3}{|c|}{ $w=-2/3$}& \multicolumn{3}{|c|}{ $w=-1$}
\\ \hline 
 $c$ & $\alpha =0.2$ & $0.4$ & $0.6$ & $0.2$ & $0.4$ & $0.6$ & $%
0.2$ & $0.4$ & $0.6$ \\ \hline
$0.001$ & $53.1411$ & $19.7498$ & $10.8993$ & $30.4544$ & $16.2782$ & $%
9.97426$ & $9.84975$ & $8.98698$ & $7.45126$ \\ 
$0.003$ & $46.5373$ & $18.3934$ & $10.3851$ & $18.0476$ & $12.0669$ & $%
8.34505$ & $5.74415$ & $5.55714$ & $5.12914$ \\ 
$0.005$ & $41.1758$ & $17.185$ & $9.91003$ & $13.4873$ & $9.84987$ & $7.26615
$ & $4.45845$ & $4.36929$ & $4.15188$ \\ 
 
 \hline
\end{tabular}
\captionof{table}{Various values of shadow radius $R_s$ for various values of $\alpha$ and $c$. Here, $M=1$ and $\ell_p=300$.} \label{table2a}
\end{center}

\begin{figure}[ht!]
    \centering
    \subfloat[$c=0.02$]{\centering{}\includegraphics[width=0.48\linewidth]{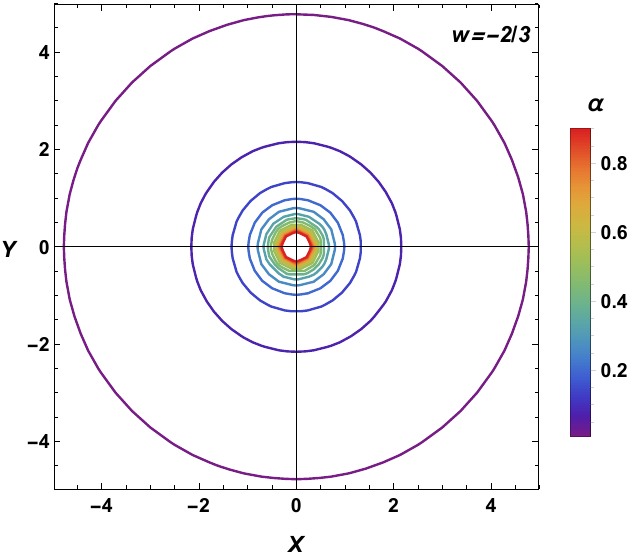}}\quad
    \subfloat[$\alpha=0.5$]{\centering{}\includegraphics[width=0.48\linewidth]{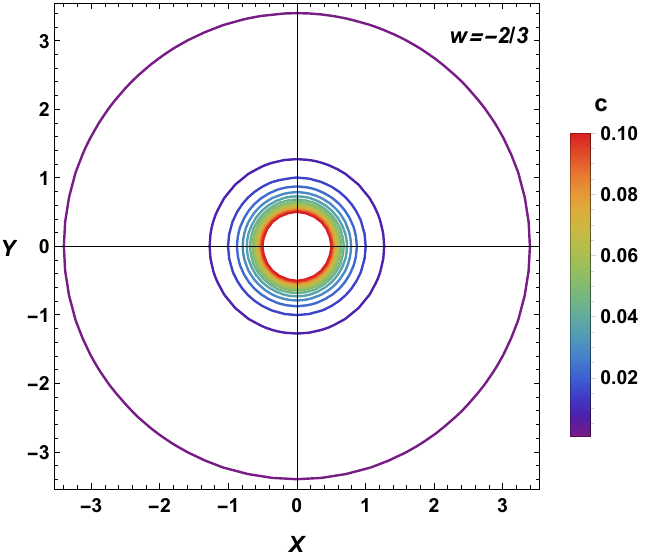}}
    \caption{The geometrical shape of the shadow radius in terms of celestial coordinates for   several values of $\alpha$ (left panel) and for $c$ (right panel). Here we set $M=1$ and $\ell_p=300$.}
    \label{figa3}
\end{figure}

 As shown in Figure \ref{figa3}, the BH shadow varies depending on the CS parameter $\alpha$ and the  positive constant of QF  parameter $c$, demonstrating the predicted dependencies in parameter space. This can be visualized in the following way: As the $\alpha$ and $c$ parameters increase, the size of the shadow shrinks. The same behavior is observed for all parameters of the QF state $w$. 
 
For circular null geodesic, we find the geodesics angular velocity given by \cite{LH1,LH4}
\begin{equation}
    \omega=\frac{\dot{\varphi}}{\dot{t}}\Big{|}_{r=r_c}=\frac{\sqrt{A(r_c)}}{r_c}=\frac{\sqrt{\alpha-\frac{2\,M}{r_c}+\frac{r^2_c}{\ell^2_{p}}+\frac{c}{r^{3\,w+1}_c}}}{r_c},\label{velocity}
\end{equation}
where we have used the relation (\ref{special2}).

From expression given in Eq. (\ref{velocity}), it becomes clear that the geodesics angular velocity is influenced by the CS parameter $\alpha$, the QF parameters $(c, w)$, the AdS radius $\ell_p$ and the BH mass $M$.

Now, we calculate the force on photon particles under the influence of the gravitational field produced by the selected BS spacetime . Various factors, such as the BH mass $M$, the CS parameter $\alpha$, the QF constant $c$, and the state parameter $w$ are responsible for the gravitational field. We show how the photon light in the given gravitational field are influenced by these factors. This force can be determined in terms of the effective potential for null geodesics by $\mathrm{F}=-V'_\text{eff}/2$, where prime denotes ordinary derivative with respect to $r$. Using the effective potential (\ref{cc1}), we find the force as follows:
\begin{equation}
    \mathrm{F}_\text{ph}=\frac{\mathrm{L}^2+\ell^2_{p}\,p^2_{z}}{r^3}\,\left(\alpha-\frac{3\,M}{r}+\frac{3\,c\,(w+1)/2}{r^{3\,w+1}}\right).\label{cc6}
\end{equation}
For a specific state parameter, for example, $w=-2/3$, from Eq. (\ref{cc6}), we find this force as, 
\begin{equation}
    \mathrm{F}_\text{ph}=\frac{\mathrm{L}^2+\ell^2_{p}\,p^2_{z}}{r^3}\,\left(\alpha-\frac{3\,M}{r}+\frac{c\,r}{2}\right).\label{cc7}
\end{equation}

The expression (\ref{cc6}) shows that the force acting on massless, light-like particles is influenced by several factors. These include the CS parameter $\alpha$, the positive constant $c$ related to the QF, the state parameter $w$, and the AdS radius $\ell_p$. Furthermore, the force is also modified by the orbital/angular momentum $\mathrm{L}$ and the translational momentum $p_z$, which together affect the force and consequently the motion of photons in the BS spacetime. 

\begin{figure}[ht!]
    \centering
    \subfloat[$\alpha=0.1$]{\centering{}\includegraphics[width=0.48\linewidth]{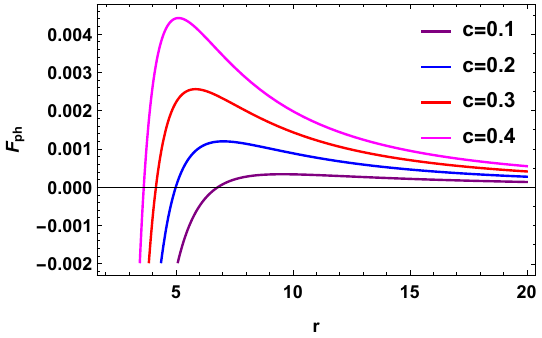}}\quad
    \subfloat[$c=0.2$]{\centering{}\includegraphics[width=0.48\linewidth]{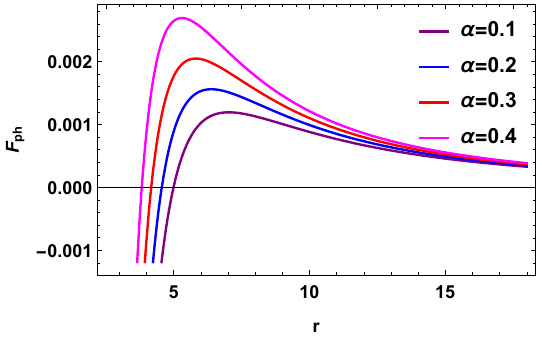}}\\
    \subfloat[]{\centering{}\includegraphics[width=0.48\linewidth]{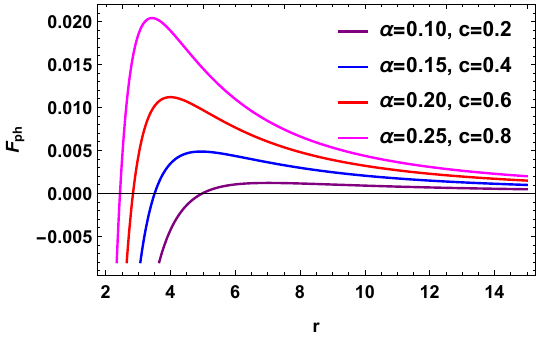}}
    \caption{Force on massless photon particles as a function of $r$. Here, $\Lambda = -0.003$, $\mathrm{L} = 1$, $p_z = 0.01$, and $M = 1$.}
    \label{fig:force}
\end{figure}

In Figure \ref{fig:force}, we illustrate the force on photon light for varying values of the CS parameter $\alpha$ and the quintessence constant $c$, with a fixed specific state parameter $w = -2/3$, cosmological constant $\Lambda = -0.003$, angular momentum $\mathrm{L} = 1$, translational momentum $p_z = 0.01$, and BH mass $M = 1$. {\color{black} It is observed that as the values of the CS parameter $\alpha$, the quintessence constant $c$, and their combination ($\alpha$, $c$) increase, the force on the photon light also increases with $r$. This suggests that larger values of these parameters increase the likelihood that photon light is captured by the BH.}

For planar null geodesics in which $p_z=0$, the force on the photon particles from Eq. (\ref{cc7}) becomes
\begin{equation}
    \mathrm{F}_\text{ph}=\frac{\mathrm{L}^2}{r^3}\,\left(\alpha-\frac{3\,M}{r}+\frac{c\,r}{2}\right).\label{cc6aa}
\end{equation}

\begin{center}
    {\bf Photon Trajectories for nonplanar geodesics}
\end{center}

Now, we focus on trajectory equation for photon particles under the influence of the gravitational field produced by a BS solution surrounded by CS and QF. Using Eqs. (\ref{bb3}), (\ref{bb4}) and (\ref{cc1}), we define the following in non-planar geodesics condition ($\dot{z} \neq 0$) as,
\begin{equation}
    \frac{\dot{r}^2}{\dot{\varphi}^2}=\left(\frac{dr}{d\varphi}\right)^2=r^4\,\left[\frac{1}{\beta^2}-\frac{\lambda^2}{\mathrm{L}^2\,r^2}\,\left(\alpha-\frac{2\,M}{r}+\frac{r^2}{\ell^2_{p}}+\frac{c}{r^{3\,w+1}}\right)\right].\label{cc8}
\end{equation}
where $\beta=\mathrm{L}/\mathrm{E}$ is the impact parameter for photon particles. 

Substituting $u=\frac{1}{r}$ into Eq. (\ref{cc8}) results in
\begin{equation}
    \left(\frac{du}{d\varphi}\right)^2+\frac{\lambda^2\,\alpha}{\mathrm{L}^2}\,u^2=\frac{1}{\beta^2}-\frac{\lambda^2}{\mathrm{L}^2\,\ell^2_{p}}+\frac{\lambda^2}{\mathrm{L}^2}\,\left(2\,M-c\,u^{3\,w}\right)\,u^3.\label{cc9}
\end{equation}

Equation (\ref{cc9}) describes the trajectory of photons in the gravitational field generated by BS solution, which includes a CS surrounded by QF. It is evident that several factors, such as the positive constant $c$ and the state parameter $w$ of QF, the CS parameter $\alpha$, the AdS radius $\ell_p$, and the momenta $(\mathrm{L}, p_z)$ all play a significant role in influencing the photon trajectory within the gravitational field of the BS solution.

We now examine the trajectory equation for a particular state parameter $w$. If $w=-2/3$, Eq. (\ref{cc9}) yields
\begin{equation}
    \left(\frac{du}{d\varphi}\right)^2+\frac{\lambda^2}{\mathrm{L}^2}\,(c\,u+\,\alpha\,u^2)=\frac{1}{\beta^2}-\frac{\lambda^2}{\mathrm{L}^2\,\ell^2_{p}}+\frac{2\,M\,\lambda^2}{\mathrm{L}^2}\,u^3.\label{cc10}
\end{equation}
Differentiating w. r. t. $\varphi$ and after simplification, we find
\begin{equation}
    \frac{d^2u}{d\varphi^2}+\frac{\lambda^2}{\mathrm{L}^2}\,\left(\frac{c}{2}+\alpha\,u\right)=\frac{3\,M\,\lambda^2}{\mathrm{L}^2}\,u^2,\quad\quad \lambda=\sqrt{\mathrm{L}^2+\ell^2_{p}\,p^2_{z}}.\label{cc11}
\end{equation}

Equation (\ref{cc11}) is a non-linear differential equation due to the presence of the quadratic term $u^2$. To obtain an analytical solution, we proceed with a perturbative approach. To this end, we assume that the solution can be expanded as a small perturbation series:
\begin{equation}
    u=u_0+\epsilon\,u_1+\epsilon^2\,u_2+...,\label{cc12}
\end{equation}
where $\epsilon$ is a small parameter that organizes the expansion.

\begin{center}
{\bf Zeroth-Order Approximation}    
\end{center}

Neglecting the non-linear term (setting $M=0$), we obtain the linear equation as follows:
\begin{equation}
    \frac{d^2u_0}{d\varphi^2}+\frac{\lambda^2\,\alpha}{\mathrm{L}^2}\,u_0=-\frac{\lambda^2\,c}{2\,\mathrm{L}^2}.\label{cc13}
\end{equation}
The homogeneous solution of the above Eq. (\ref{cc13}) is given by
\begin{equation}
    u^{(h)}_0=a_1\,\cos(k\,\varphi)+a_2\,\sin(k\,\varphi),\quad\quad k^2=\frac{\lambda^2\,\alpha}{\mathrm{L}^2},\label{cc14}
\end{equation}
where $a_1$ and $a_2$ are arbitrary constants. On the other hand, a particular solution for the inhomogeneous Eq. (\ref{cc13}) is given by
\begin{equation}
    u^{(p)}_0=-\frac{c\,k^2}{2\,\alpha}.\label{cc15}
\end{equation}

Thus, the zeroth-order solution for the Eq. (\ref{cc11}) is given by
\begin{equation}
    u_0= u^{(h)}_0+ u^{(p)}_0=a_1\,\cos(k\,\varphi)+a_2\,\sin(k\,\varphi)-\frac{c\,k^2}{2\,\alpha}.\label{cc16}
\end{equation} 

\begin{center}
{\bf First-Order Approximation}    
\end{center}

For the first-order solution for $u$, Eq. (\ref{cc11}) can be written as:
\begin{equation}
    \frac{d^2u_1}{d\varphi^2}+k^2\,u_1=\frac{3\,M\,\lambda^2}{\mathrm{L}^2}\,u^2_{0}=\frac{3\,M\,\lambda^2}{\mathrm{L}^2}\,\left(a_1\,\cos(k\,\varphi)+a_2\,\sin(k\,\varphi)-\frac{c\,k^2}{2\,\alpha}\right)^2.\label{cc17}
\end{equation}

Therefore, the right-hand side of Eq. (\ref{cc17}) can be written as:
\begin{equation}
    \frac{3\,M\,\lambda^2}{\mathrm{L}^2}\,\left(a_1\,\cos(k\,\varphi)+a_2\,\sin(k\,\varphi)-\frac{c\,k^2}{2\,\alpha}\right)^2=a_3+a_4\,\cos (2\,k\,\varphi)+a_5\,\sin (2\,k\,\varphi)+a_6\,\cos(k\,\varphi)+a_7\,\sin (k\,\varphi),\label{cc18}
\end{equation}
where the coefficients $a_3, a_4 \cdots, a_7$ are related with the mass $M$, angular momentum $\mathrm{L}$, translational momentum $p_z$, CS parameter $\alpha$, QF constant $c$ for a specific state parameter $w=-2/3$, and the radius of curvature $\ell_p$.

Th complementary solution to the homogeneous Eq.(\ref{cc17}) is given by
\begin{equation}
    u^{(h)}_1=a_8\,\cos(k\,\varphi)+a_9\,\sin(k\,\varphi),\label{cc19}
\end{equation}
where $a_8$ and $a_9$ are arbitrary constants.

Using Eq. (\ref{cc18}) into the above nonlinear Eq. (\ref{cc17}), we find the particular solution for $u_1$ as follows:
\begin{equation}
    u^{(p)}_1=a_{10}+a_{11}\,\cos(2\,k\,\varphi)+a_{12}\,\sin(2\,k\,\varphi),\label{cc20}
\end{equation}
where the coefficients $a_{10}$, $a_{11}$ and $a_{12}$ are related with $a_3, a_4 \cdots, a_7$ stated earlier.

The general solution for $u_1$ is the sum of the homogeneous solution (\ref{cc19}) and the particular solution (\ref{cc20}). Thus, we find
the general form for $u_1$ as:
\begin{equation}
    u_1=a_8\,\cos(k\,\varphi)+a_9\,\sin(k\,\varphi)+a_{10}+a_{11}\,\cos(2\,k\,\varphi)+a_{12}\,\sin(2\,k\,\varphi).\label{cc21}
\end{equation}

Hence, the general solution of the non-linear differential Eq. (\ref{cc11}) up to the first-order approximation is given by
\begin{eqnarray}
    \frac{1}{r(\varphi)}=u(\varphi)&=&-\frac{c\,k^2}{2\,\alpha}+a_1\,\cos(k\,\varphi)+a_2\,\sin(k\,\varphi)+\epsilon\,\Big[a_8\,\cos(k\,\varphi)+a_9\,\sin(k\,\varphi)+a_{11}\,\cos(2\,k\,\varphi)+a_{12}\,\sin(2\,k\,\varphi)\Big]\nonumber\\
    &=&\mathcal{C}_0+\mathcal{C}_1\,\cos(k\,\varphi)+\mathcal{C}_2\,\sin (k\,\varphi)+\mathcal{C}_3\,\cos \left(2\,k\,\varphi\right)+\mathcal{C}_4\,\sin (2\,k\,\varphi),\nonumber\\
    k&=&\sqrt{\alpha\,\left(1+\frac{\ell^2_p}{\mathrm{L}^2}\,\,p^2_z\right)}=\sqrt{\alpha\,\left(1-\frac{3\,p^2_z}{\Lambda\,\mathrm{L}^2}\right)}.\label{solution}
\end{eqnarray}
where $\mathcal{C}_0,\cdots, \mathcal{C}_4$ are related with the BH mass $M$, the angular momentum $\mathrm{L}$, the translational momentum $p_z$, the CS parameter $\alpha$, the QF constant $c$ for a specific state parameter $w=-2/3$, and the radius of curvature $\ell_p$.

From expression provided in Eq. (\ref{solution}), it becomes clear that the trajectory of the geodesic path $r(\varphi)$ is influenced by a variety of factors, each contributing to its behavior in distinct ways. Key among these is the CS parameter $\alpha$, which plays a significant role in determining the curvature and geometry of spacetime. Additionally, the QF constant $c$ for a specific equation of state parameter $w = -2/3$, which is often associated with a scalar field model of dark energy, also has a profound impact on the geodesic's path. This constant modifies the energy-momentum distribution in the spacetime, affecting the overall structure and evolution of the geodesic.

Another critical factor is the radius of curvature $\ell_p$, which is related to the size and shape of the spacetime under consideration. The radius of curvature dictates how spacetime itself bends, influencing the geodesic trajectory, particularly in non-flat geometries such as those encountered in cosmological models.

Furthermore, both the angular momentum $\mathrm{L}$ and the translational momentum $p_z$ play pivotal roles in shaping the geodesic path. The angular momentum $\mathrm{L}$ governs the rotational characteristics of the path, determining the amount of "twist" or curvature experienced by the object as it moves along the geodesic. Similarly, the translational momentum $p_z$, related to the linear motion along the $z$-axis, impacts the geodesic's trajectory in the vertical direction, influencing the overall movement of the particle or object in question.

Taken together, these factors-$\alpha$, $c$, $\ell_p$, $\mathrm{L}$, and $p_z$-interact in complex ways to modify the shape and characteristics of the geodesic path. Each one contributes to the overall dynamical evolution, and their interplay defines the motion of objects within this spacetime framework. Thus, understanding how these parameters affect the geodesics is essential for a comprehensive analysis of the underlying spacetime geometry and the forces that govern the motion of particles or light within it.

{\color{black}

\begin{center}
    {\bf Photon Trajectories for planar geodesics}
\end{center}

Now, we focus on photon trajectories under planar geodesics condition in which $p_z=0$, and hence, $\lambda^2=\mathrm{L}^2$. The trajectory equation for photon particles from (\ref{cc11}) therefore reduces as
\begin{equation}
    \frac{d^2u}{d\varphi^2}+\alpha\,u=3\,M\,u^2-\frac{c}{2}.\label{nn2}
\end{equation}

Following a similar procedure performed earlier, one can determine the general solution of Eq. (\ref{nn2}) up to the first-order approximation given by (here $k^2=\alpha$ using Eq. (\ref{cc14}) since $\lambda^2=\mathrm{L}^2$)
\begin{eqnarray}
    \frac{1}{r(\varphi)}=u(\varphi)&=&-\frac{c}{2}+a_1\,\cos(\sqrt{\alpha}\,\varphi)+a_2\,\sin(\sqrt{\alpha}\,\varphi)+\epsilon\,\Big[a_8\,\cos(\sqrt{\alpha}\,\varphi)+a_9\,\sin(\sqrt{\alpha}\,\varphi)+a_{11}\,\cos(2\,\sqrt{\alpha}\,\varphi)\nonumber\\
    &+&a_{12}\,\sin(2\,\sqrt{\alpha}\,\varphi)\Big]\nonumber\\
    &=&-\frac{c}{2}+\mathcal{D}_1\,\cos(\sqrt{\alpha}\,\varphi)+\mathcal{D}_2\,\sin (\sqrt{\alpha}\,\varphi)+\mathcal{D}_3\,\cos \left(2\,\sqrt{\alpha}\,\varphi\right)+\mathcal{D}_4\,\sin (2\,\sqrt{\alpha}\,\varphi),\label{nn3}
\end{eqnarray}
where $\mathcal{D}_1,\cdots, \mathcal{D}_4$ are related with the BH mass $M$ and the QF normalization constant $c$ for a specific state parameter $w=-2/3$.

From expression provided in Eq. (\ref{nn3}), it becomes clear that photon trajectory $r(\varphi)$ in planar geodesic condition is influenced by the CS parameter $\alpha$ and the QF normalization constant $c$ for a specific state parameter $w = -2/3$.
}

\subsubsection{\bf Time-like geodesics}

{\color{black} A time-like geodesic is the path that a massive particle follows under the influence of gravity alone. Unlike null geodesics, which describe the path of light or massless particles, time-like geodesics are associated with particles that have a non-zero rest mass, such as planets or other massive objects. The effective potential for time-like geodesics is a tool used to study the motion of massive particles in the curved spacetime around a BH. It simplifies the equations of motion, particularly when studying radial motion (motion along the radial direction). The effective potential helps determine the particle's behavior as it interacts with BH's gravity, including whether the particle can escape, orbit, or fall into the BH's event horizon.}

For massive objects $\varepsilon=-1$, the effective potential for time-like geodesics from (\ref{bb5}) becomes
\begin{equation}
    V_\text{eff}=\left(1+\frac{\lambda^2}{r^2}\right)\,\left(\alpha-\frac{2\,M}{r}+\frac{r^2}{\ell^2_p}+\frac{c}{r^{3\,w+1}}\right).\label{dd1}
\end{equation}
And the geodesics equation for the radial coordinate $r$ is given by
\begin{equation}
    \dot{r}=\sqrt{\mathrm{E}^2-V_\text{eff}},\label{dd2}
\end{equation}
where $V_\text{eff}$ is given in Eq. (\ref{dd1}). 

\begin{figure}[ht!]
    \centering
    \subfloat[$\alpha=0.1$]{\centering{}\includegraphics[width=0.48\linewidth]{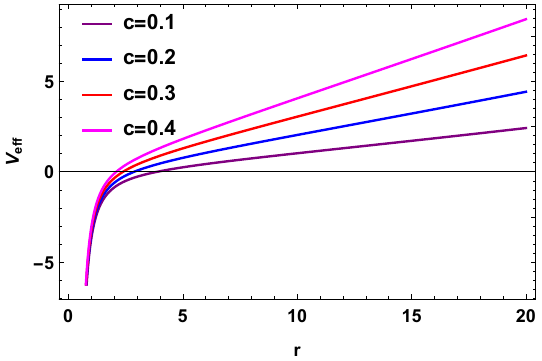}}\quad
    \subfloat[$c=0.2$]{\centering{}\includegraphics[width=0.48\linewidth]{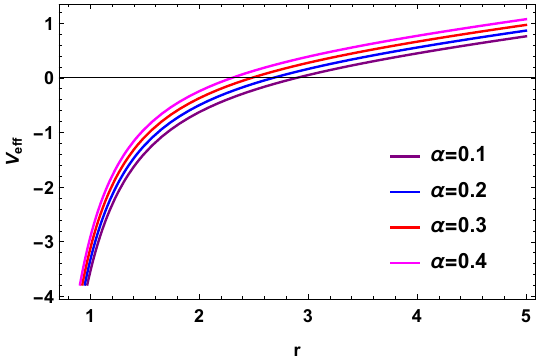}}\\
    \subfloat[]{\centering{}\includegraphics[width=0.48\linewidth]{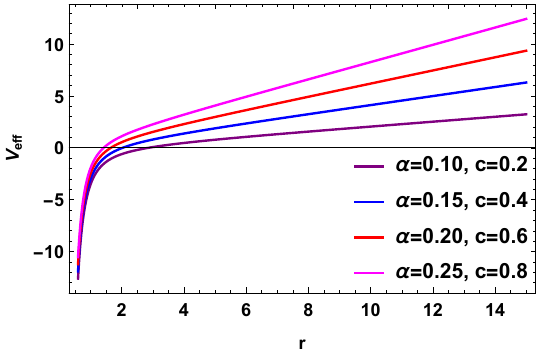}}
    \caption{The effective potential for non-planar time-like geodesics, $\dot{z} \neq 0$. Here, $\Lambda = -0.003$, $\mathrm{L} = 1$, $\mathrm{p}_z = 0.01$, and $M=1$.}
    \label{fig:time-like}
\end{figure}

In Figure \ref{fig:time-like}, we present the effective potential for time-like geodesics given in Eq. (\ref{dd1}) with varying values of the CS parameter $\alpha$ and the quintessence constant $c$, with a fixed specific state parameter $w = -2/3$,  cosmological constant $\Lambda = -0.003$, angular momentum $\mathrm{L} = 1$, translational momentum $\mathrm{p}_z = 0.01$, and BH mass $M = 1$. {\color{black} It is observed that as the values of CS parameter $\alpha$, the quintessence constant $c$, and their combination ($\alpha, c$) increase, the effective potential for time-like particle also increases with increasing $r$. This suggests that the massive object is more strongly attracted by the BH's gravity as the values of these parameters increase.}

Now, we determine the energy and angular momentum for time-like particles orbiting the circular time-like geodesics ($\dot{z}=0$) around the BH. For circular helices along cylindrical surfaces or planar circles, we employ the conditions $\dot{r}=0$ and $\ddot{r}=0$ which yields 
\begin{eqnarray}
    \mathrm{E}^2=V_\text{eff}=\left(1+\frac{\mathrm{L}^2}{r^2}\right)\,\left(\alpha-\frac{2\,M}{r}+\frac{r^2}{\ell^2_p}+\frac{c}{r^{3\,w+1}}\right)\quad \mbox{and} \quad V'_\text{eff}(r)=0,\label{dd0}
\end{eqnarray}
where prime denotes ordinary derivative w. r. t. $r$.

By simplifying the aforementioned relations, we find the radial dependent angular momentum associated with time-like particles expressed as
\begin{equation}
    \mathrm{L}(r)=r\,\sqrt{\frac{\frac{M}{r}+\frac{r^2}{\ell^2_{p}}-\frac{c\,(3\,w+1)/2}{r^{3\,w+1}}}{\alpha-\frac{3\,M}{r}+\frac{3\,c\,(w+1)/2}{r^{3\,w+1}}}},\label{dd3}
\end{equation}
and the particles' energy reads
\begin{equation}
    \mathrm{E}_{\pm}=\pm\,\frac{\left(\alpha-\frac{2\,M}{r}+\frac{r^2}{\ell^2_{p}}+\frac{c}{r^{3\,w+1}}\right)}{\sqrt{\alpha-\frac{3\,M}{r}+\frac{3\,c\,(w+1)/2}{r^{3\,w+1}}}}.\label{dd4}
\end{equation}
From the above particles' energy expression, we see that the energy approaches $\mathrm{E}_{\pm} \to \pm\,\sqrt{\alpha}$ as $r \to \infty$ and $\ell_p \to \infty$. Hence, the maximum particles' energy will be $E^{\max}=\sqrt{\alpha}$ which is less than unity (1).

From the angular momentum and energy expressions given in Eqs. (\ref{dd3}) and (\ref{dd4}), it is evident that various factors, such as the positive constant $c$ and state parameter $w$ of QF, the CS parameter $\alpha$, and the AdS radius $\ell_p$, all play a significant role in influencing the particles motion in circular orbits within the gravitational field of the BS solution. Additionally, the BH mass $M$ also influences these physical quantities.

For a specific state parameter, $w=-2/3$, the physical quantities given in Eqs. (\ref{dd3}) and (\ref{dd4}) reduce as:
\begin{equation}
    \mathrm{L}(r)=r\,\sqrt{\frac{\frac{M}{r}+\frac{r^2}{\ell^2_{p}}+\frac{c\,r}{2}}{\alpha-\frac{3\,M}{r}+\frac{c\,r}{2}}},\quad\quad \mathrm{E}_{\pm}=\pm\,\frac{\left(\alpha-\frac{2\,M}{r}+\frac{r^2}{\ell^2_{p}}+c\,r\right)}{\sqrt{\alpha-\frac{3\,M}{r}+\frac{c\,r}{2}}},\quad\quad \frac{1}{\ell^2_p}=-\frac{\Lambda}{3}.\label{case}
\end{equation}

\begin{figure}[ht!]
    \centering
    \subfloat[$\alpha=0.1$]{\centering{}\includegraphics[width=0.48\linewidth]{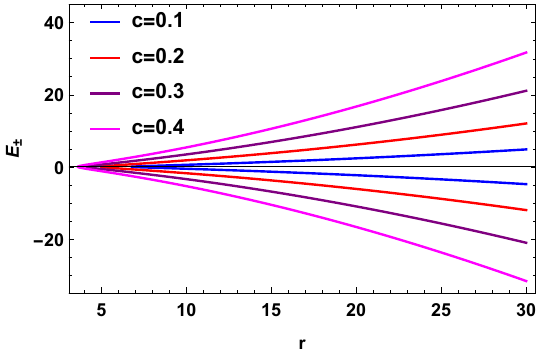}}\quad
    \subfloat[$c=0.2$]{\centering{}\includegraphics[width=0.48\linewidth]{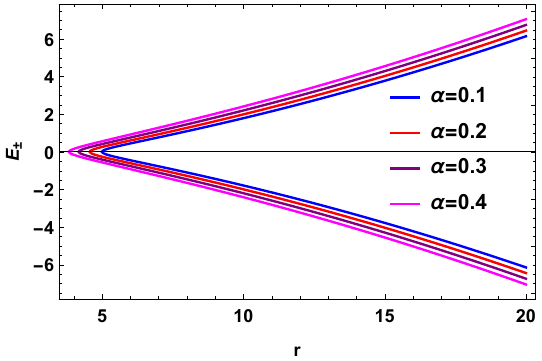}}\\
    \subfloat[]{\centering{}\includegraphics[width=0.48\linewidth]{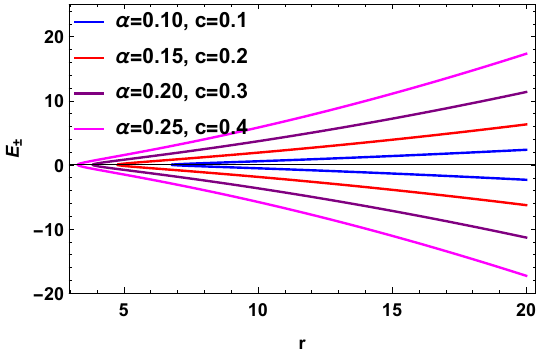}}
    \caption{The energy of time-like particles in non-planar geodesics, $\dot{z} \neq 0$. Here, $\Lambda = -0.003$, $\mathrm{L} = 1$, $\mathrm{p}_z = 0.01$, and $M=1$.}
    \label{fig:energy}
\end{figure}

In Figure \ref{fig:energy}, we show the energy of time-like particles orbiting in circular geodesics. The plot varies the values of the CS parameter $\alpha$ and the quintessence constant $c$, with a fixed specific state parameter $w = -2/3$, cosmological constant $\Lambda = -0.003$, angular momentum $\mathrm{L} = 1$, translational momentum $\mathrm{p}_z = 0.01$, and BH mass $M = 1$. {\color{black} It is observed that as the values of the CS parameter $\alpha$ and the quintessence constant $c$ increase, the particles' energy also increases with increasing $r$. This indicates that higher values of these parameters result in more energy for time-like particles, enhancing their tendency to orbit in  circular geodesics around the BH.}

Moreover, the angular velocity of the test particles on circular orbits is defined by \cite{VC}
\begin{eqnarray}
    \Omega=\frac{\dot{\varphi}}{\dot{t}}=\frac{A(r)}{r^2}\,\frac{\mathrm{L}}{\mathrm{E}}=\sqrt{\frac{A'(r)}{2\,r}}=\sqrt{\frac{M}{r^3}+\frac{1}{\ell^2_{p}}-\frac{c\,(3\,w+1)/2}{r^{3\,(w+1)}}}.\label{dd5}
\end{eqnarray}
The expression (\ref{dd5}) shows that the angular velocity of time-like particles in circular orbits is affected by the CS parameter $\alpha$, the positive constant $c$ of the QF for a specific state parameter $w$, and the AdS radius $\ell_p$. 

For a specific state parameter, for example, $w=-2/3$, the angular velocity from Eq. (\ref{dd5}) becomes
\begin{eqnarray}
    \Omega=\sqrt{\frac{A'(r)}{2\,r}}=\sqrt{\frac{M}{r^3}+\frac{1}{\ell^2_{p}}+\frac{c}{2\,r}}.\label{dd5aa}
\end{eqnarray}

\subsection{\large \bf The C-energy of AdS BS solution}\label{sec:4}

{\color{black} The C-energy for a cylindrical BH represents the conserved energy of a particle or object moving in the spacetime around the BH. It reflects how the energy of the particle changes as it moves through regions of varying gravitational influence, and in cylindrical BHs, the C-energy plays a crucial role in understanding the particle dynamics, orbital behavior, and accretion processes.}

For spacetime (\ref{aa1}), the metric tensor is independent of the coordinates ($\varphi,z$). Thus, there exist Killing vectors that are $\xi_{(\varphi)} \equiv \partial_{\varphi}$ associated with invariant rotations and $\xi_{(z)} \equiv \partial_{\varphi}$ associated with the invariant translations of the cylindrical system. The existence of axial symmetry about the $z$-axis (since $|g_{\varphi\varphi}| \to 0$, as $r \to 0$) implies that the orbits of one of these Killing vectors, namely $\partial_{\varphi}$, are closed but those of the other $\partial_{z}$ are open.

To normalize the Killing vector $\xi^{\,\mu}_{(z)}$ within the metric, we find
\begin{equation}
    |\xi_{(z)}|=\sqrt{\xi^{\,\,\mu}_{(z)}\,\xi_{(z){\,\mu}}}=1,\quad \xi^{\,\mu}_{(z)}=\left(0,0,0,\frac{\ell_p}{r}\right),\quad \xi_{(z){\,\mu}}=\left(0,0,0,\frac{r}{\ell_p}\right)\label{ss1}
\end{equation}
along the symmetry axis when gravitational radiation is absent, as shown in \cite{KST1,KST2}.

The normalization condition imposed on the Killing vector $\xi^{\,\mu}_{(\varphi)}$ for our metric is defined by
\begin{eqnarray}
    &&|\xi_{(\varphi)}|=\sqrt{\xi^{\,\,\mu}_{(\varphi)}\xi_{(\varphi){\,\mu}}}=2\pi r\,(\text{the proper circumference around the symmetry axis}),\nonumber\\
 &&\xi^{\,\,\mu}_{(\varphi)}=\left(0,0,2\,\pi,0\right),\,\, \xi_{(\varphi){\,\mu}}=\left(0,0,2\,\pi\,r^2,0\right).\label{ss2}
\end{eqnarray}

Examining the system's invariant transformations, we have:
\begin{equation}
    x^{\mu} \to x^{\mu} + \lambda\,\xi^{\,\,\mu}_{(z)} + \eta\,\xi^{\,\,\mu}_{(\varphi)},\quad 0 \leq \lambda \leq \lambda_0,\quad 0 \leq \eta \leq 1.\label{ss3}
\end{equation}

A specific point \( P \) in spacetime traces a cylindrical path with length
\begin{equation}
    Z=\lambda_0\,|\xi_{(z)}|,\label{ss4}
\end{equation}
and area
\begin{equation}
    \mathcal{A}=Z\,|\xi_{(\varphi)}|=\lambda_0\,|\xi_{(z)}||\xi_{(\varphi)}|=2\,\pi\,r\,\lambda_0.\label{ss5}
\end{equation}

The potential function for C energy is defined by \cite{KST1,KST2}
\begin{equation}
    U=-\frac{1}{4}\,\ln\left\{\frac{\sqrt{g^{\mu\nu}\,\mathcal{A}_{,\mu}\,\mathcal{A}_{,\nu}}}{2\,\pi\,Z}\right\}=-\frac{1}{8}\,\ln\left\{\frac{g^{\mu\nu}\,\mathcal{A}_{,\mu}\,\mathcal{A}_{,\nu}}{4\,\pi^2\,Z^2}\right\}=-\frac{1}{8}\,\ln\left\{\frac{g^{\mu\nu}\,\left(\mathcal{A}_{,\mu}/\lambda_0\right)\,\left(\mathcal{A}_{,\nu}/\lambda_0\right)}{4\,\pi^2\,|\xi_{(z)}|^2} \right\}=-\frac{1}{8}\,\ln (A(r)).\label{ss6}
\end{equation}

In terms of the geometrically defined scalar field $U$, and the Killing vectors $\xi_{(z)}, \xi_{(\varphi)}$ for whole-cylinder
symmetry, the C-energy is defined in terms of the C-energy flux vector $P^{\mu}$ which satisfies the conservation law $\nabla_{\mu}\,P^{\mu}=P^{\mu}_{\,\,;\mu}=0$. The C-energy flux vector $P^{\mu}$ is defined by \cite{KST1,KST2}
\begin{equation}
    P^{\mu}=\frac{\epsilon^{\mu\nu\rho\tau}}{\sqrt{-g}}\,\frac{U_{,\nu}\,\xi_{(z){\,\rho}}\,\xi_{(\varphi){\,\tau}}}{|\xi_{(z)}|^2\,|\xi_{(\varphi)}|^2}=\frac{A'(r)}{16\,\pi\,r\,A(r)}\,(1,0,0,0)\label{ss7}
\end{equation}
where $\epsilon^{\mu\nu\rho\tau}$ is the 4-dimensional skew tensor.

Therefore, the C-energy density measured by an observer whose worldline has a unit tangent $u^{\mu}$ that satisfies the normalization condition $u^{\mu}\,u_{\mu}=-1$. In our case, we define time-like and spacelike normalized unit vectors as,
\begin{eqnarray}
    &&u^{\mu}=\frac{1}{\sqrt{A(r)}}\,(1,0,0,0),\quad u_{\mu}=\sqrt{A(r)}\,(-1,0,0,0),\quad u^{\mu}\,u_{\mu}=-1,\nonumber\\
    &&n^{\mu}=\sqrt{A(r)}\,(0,1,0,0),\quad n_{\mu}=\frac{1}{\sqrt{A(r)}}\,(0,1,0,0),\quad n^{\mu}\,n_{\mu}=1,\quad
    u^{\mu}\,n_{\mu}=0.\label{ss8}
\end{eqnarray}

Thus, the C-energy density measured by an observer in a local Lorentz frame is given by \cite{KST1,KST2}
\begin{equation}
    \mathcal{E}_{ce}=-P^{\mu}\,u_{\mu}=\frac{A'(r)}{16\,\pi\,r\,\sqrt{A(r)}}=\frac{1}{16\,\pi\,r^2}\,\frac{\frac{2\,M}{r}+\frac{2\,r^2}{\ell^2_{p}}-\frac{c\,(3\,w+1)}{r^{3\,w+1}}}{\sqrt{\alpha-\frac{2\,M}{r}+\frac{r^2}{\ell^2_{p}}+\frac{c}{r^{3\,w+1}}}}.\label{ss9}
\end{equation}
The C-energy flux which the observer sees flowing in a direction $n^{\mu}$ orthogonal to his/her world line is
\begin{equation}
    \mathcal{F}_{ce}=P^{\mu}\,n_{\mu}=0.\label{ss10}
\end{equation}

From Eq. (\ref{ss9}), it is clear that several factors-such as the positive constant $c$ and the state parameter $w$ of QF, the CS parameter $\alpha$, and the AdS radius $\ell_p$-play a significant role in determining the density of the C energy. This density approaches zero, $\mathcal{E}_{ce} \to 0$ as $r \to \infty$.

When $w=-2/3$, the C-energy density is given by
\begin{equation}
    \mathcal{E}_{ce}=\frac{1}{16\,\pi\,r^2}\,\frac{\left(\frac{2\,M}{r}+\frac{2\,r^2}{\ell^2_{p}}+c\,r\right)}{\sqrt{\alpha-\frac{2\,M}{r}+\frac{r^2}{\ell^2_{p}}+c\,r}},\quad\quad \frac{1}{ell^2_p}=-\frac{\Lambda}{3}.\label{ss11}
\end{equation}
Conversely, when $w=-1/3$, it is expressed as
\begin{equation}
    \mathcal{E}_{ce}=\frac{1}{16\,\pi\,r^2}\,\frac{\frac{2\,M}{r}+\frac{2\,r^2}{\ell^2_{p}}}{\sqrt{\alpha-\frac{2\,M}{r}+\frac{r^2}{\ell^2_{p}}+c}}.\label{ss12}
\end{equation}

\begin{figure}[ht!]
    \centering
    \subfloat[$\alpha=0.1$]{\centering{}\includegraphics[width=0.48\linewidth]{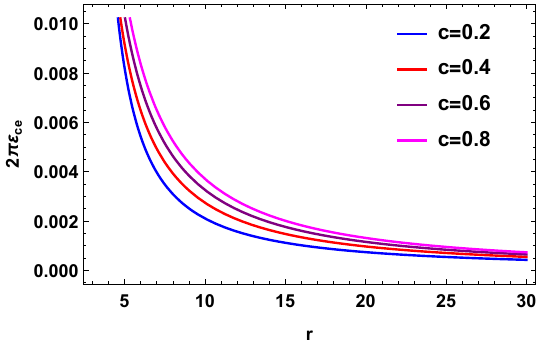}}\quad
    \subfloat[$c=0.2$]{\centering{}\includegraphics[width=0.48\linewidth]{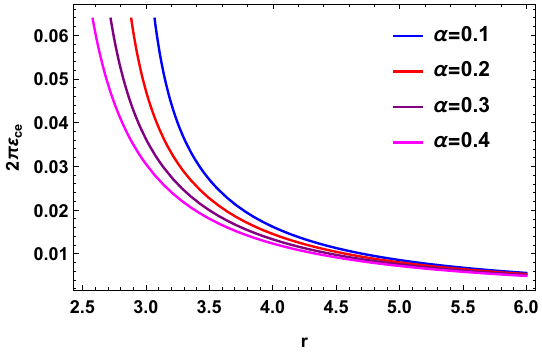}}\\
    \subfloat[]{\centering{}\includegraphics[width=0.48\linewidth]{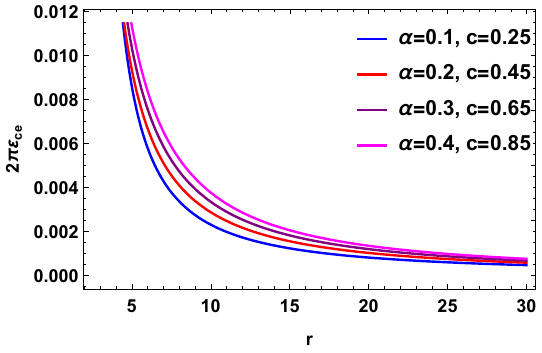}}
    \caption{The C-energy density as a function of $r$ for BH. Here $M=1$ and $\Lambda = -0.003$.}
    \label{fig:C-Energy}
\end{figure}

In Figure \ref{fig:C-Energy}, we plot the C-energy density with varying the values of the CS parameter $\alpha$ and the quintessence constant $c$, with a fixed specific state parameter $w = -2/3$, cosmological constant $\Lambda = -0.003$, and BH mass $M = 1$. {\color{black} It is observed that for specific values of the CS parameter $\alpha$, the quintessence constant $c$, and their combination $(\alpha, c)$, the C-energy gradually decreases as the radial distance $r$ increases. As the C-energy decreases with $r$, the likelihood of a particle being captured by the BH increases, especially as the energy approaches zero near the event horizon ($r_{+}$) or other critical regions. This behavior is crucial for understanding the accretion dynamics near BHs.

In Figure \ref{fig:C-Energy}, panel (a) shows that the decreasing trend of C-energy shifts upward as the value of the QF constant $c$ increases. In contrast, panel (b) shows that this trend shifts downward as the value of the CS parameter $\alpha$ increases. Finally, in panel (c), the decreasing trend shifts upward again when both the CS parameter $\alpha$ and the QF constant $c$ are increased together.
}

\subsection{\large \bf Scalar Perturbations for AdS BS Spacetime: The massless Klein-Gordon Equation} \label{sec:5}

In this section, we examine the dynamics of a massless scalar field in the background of a selected AdS black string solution and show how various parameters involved in the spacetime geometry alters the perturbative potential. We begin by deriving the massless Klein-Gordon equation that governs the evolution of the scalar field in the AdS BS spacetime geometry. Scalar perturbations are critical for understanding BH stability, particularly in the context of BH spacetimes. These perturbations have been extensively studied across various BH solutions in GR, providing valuable insights into BH stability and the propagation of fields in curved spacetime. For example, scalar field perturbations have been analyzed in Schwarzschild, Kerr, and Reissner-Nordstr\"{o}m BHs, as well as in other BH solutions in both GR and modified gravity theories (see, e.g., Refs. \cite{NPB,CJPHY,AHEP1,AHEP2,EPJC}, and related works).

The massless scalar field wave equation is described by the Klein-Gordon equation as follows \cite{NPB,CJPHY,AHEP1,AHEP2,EPJC}:
\begin{equation}
\frac{1}{\sqrt{-g}}\,\partial_{\mu}\left(\sqrt{-g}\,g^{\mu\nu}\,\partial_{\nu}\Psi\right)=0\quad\quad (\mu,\nu=0,\cdots,3),\label{ff1}    
\end{equation}
where $\Psi$ is the wave function of the scalar field, $g_{\mu\nu}$ is the covariant metric tensor, $g=\det(g_{\mu\nu})$ is the determinant of the metric tensor, $g^{\mu\nu}$ is the contrvariant form of the metric tensor, and $\partial_{\mu}$ is the partial derivative with respect to the coordinate systems.

For the AdS BS spacetime (\ref{aa1}), we have labeled the coordinates $x^0=t$, $x^1=r$, $x^2=\varphi$ and $x^3=z$. Moreover, we find the following for the given spacetime as, {\small
\begin{eqnarray}
    g_{\mu\nu}=\mbox{diag}\left(-A(r),\,\frac{1}{A(r)},\, r^2,\, \frac{r^2}{\ell^2_{p}}\right),\quad
    g^{\mu\nu}=\mbox{diag}\left(-\frac{1}{A(r)},\,A(r),\, \frac{1}{r^2},\, \frac{\ell^2_{p}}{r^2}\right),\quad
    g=\mbox{det} (g_{\mu\nu})=-\frac{r^4}{\ell^2_{p}},\quad \sqrt{-g}=\frac{r^2}{\ell_{p}}.\label{ff0}
\end{eqnarray}
}
With these, we can write Eq. (\ref{ff1}) explicitly as follows:
\begin{eqnarray}
    -\frac{1}{A(r)}\,\partial^2_{t}\,\Psi({\bf r})+\frac{1}{r^2}\,\partial_r\,\left(r^2\,A\,\partial_r\,\Psi({\bf r})\right)+\frac{1}{r^2}\,\partial^2_{\varphi}\,\Psi({\bf r})+\frac{\ell^2_{p}}{r^2}\,\partial^2_{z}\,\Psi({\bf r})=0.\label{ff}
\end{eqnarray}

Since the selected spacetime is a static and cylindrically symmetric, let us consider the following scalar field ansatz form
\begin{equation}
    \Psi=\Psi(t, r, \varphi, z)=\exp(i\,\omega\,t)\,\exp(i\,m\,\varphi)\,\exp(i\,k\,z)\,\frac{\psi(r)}{r},\label{ff2}
\end{equation}
where $\omega$ is (possibly complex) the temporal frequency, $\psi (r)$ is a propagating scalar field in the candidate spacetime, $m$ is the multipole quantum number and $k$ is an arbitrary constant.

Substituting the scalar field ansatz Eq. (\ref{ff2}) into the Eq. (\ref{ff}), we find the following second-order differential equation:
\begin{equation}
    A^2(r)\,\psi''(r)+A(r)\,A'(r)\,\psi'(r)+\left(\omega^2-\frac{\sigma^2}{r^2}\,A(r)\right)\,\psi(r)=0,\quad\quad \sigma=\sqrt{m^2+k^2\,\ell^2_{p}}.\label{ff3}
\end{equation}

Performing the following coordinate change (called tortoise coordinate) 
\begin{eqnarray}
    r_*=\int\,\frac{dr}{A(r)},\quad\quad \partial_{r_*}=A(r)\,\partial_{r}.\label{ff4}
\end{eqnarray}

One can rewrite the wave Eq. (\ref{ff3}) in the following form:
\begin{equation}
    \frac{\partial^2 \psi(r_*)}{\partial r^2_{*}}+\left(\omega^2-\mathcal{V}\right)\,\psi(r_*)=0,\label{ff5}
\end{equation}
which is equivalent to the time-independent Schrodinger-like wave equation. The perturbative scalar potential $\mathcal{V}$ is given as,
\begin{eqnarray}
\mathcal{V}&=&\left(\frac{\sigma^2}{r^2}+\frac{A'(r)}{r}\right)\,A(r).\label{ff6}
\end{eqnarray}

Substituting the metric function $A(r)$ into the Eq. (\ref{ff6}), we find following expression of the scalar perturbative potential
\begin{eqnarray}
\mathcal{V}_m=\left(\frac{m^2+k^2\,\ell^2_{p}}{r^2}+\frac{2\,M}{r^3}+\frac{2}{\ell^2_{p}}-\frac{c\,(3\,w+1)}{r^{3\,(w+1)}}\right)\,\left(\alpha-\frac{2\,M}{r}+\frac{r^2}{\ell^2_{p}}+\frac{c}{r^{3\,w+1}}\right).\label{ff7}
\end{eqnarray}

{\color{black} From expression in Eq. (\ref{ff7}),we observe that various factors involved in the spacetime geometry as well as others influences this perturbative potential. These include the CS parameter $\alpha$,the QF parameter $(c, w)$, the AdS radius $\ell_p$, and the BH mass $M$. Additionally, the multipole quantum number $m$ and the constant $k$ also modified this scalar potential.
}

As for example, we discuss the above perturbative scalar potential for a particular state parameter, $w=-2/3$. Therefore, from Eq. (\ref{ff7}), we find
\begin{eqnarray}
\mathcal{V}_m=\left(\frac{m^2+k^2\,\ell^2_{p}}{r^2}+\frac{2\,M}{r^3}+\frac{2}{\ell^2_{p}}+\frac{c}{r}\right)\,\left(\alpha-\frac{2\,M}{r}+\frac{r^2}{\ell^2_{p}}+c\,r\right).\label{ff8}
\end{eqnarray}

Defining dimensionless variables $x=r/M$, $y=c\,M$ and $\mathrm{a}=M/\ell_p$, we define the following quantity
\begin{eqnarray}
M^2\,\mathcal{V}_m=\left(\frac{\sigma^2}{x^2}+\frac{2}{x^3}+2\,\mathrm{a}^2+\frac{y}{x}\right)\,\left(\alpha-\frac{2}{x}+\mathrm{a}^2\,x^2+x\,y\right).\label{ff9}
\end{eqnarray}

In the limit where $\ell_p \to \infty$ which implies $\mathrm{a} \to 0$, no AdS background, the above dimensionless quantity becomes
\begin{eqnarray}
M^2\,\mathcal{V}_m=\frac{\left[x^2\,y+(m^2+k^2\,\ell^2_{p})\,x+2\right]\left(x^2\,y+\alpha\,x-2\right)}{x^4}.\label{ff10}
\end{eqnarray}

\begin{figure}[ht!]
    \centering
    \subfloat[$m=0$]{\centering{}\includegraphics[width=0.48\linewidth]{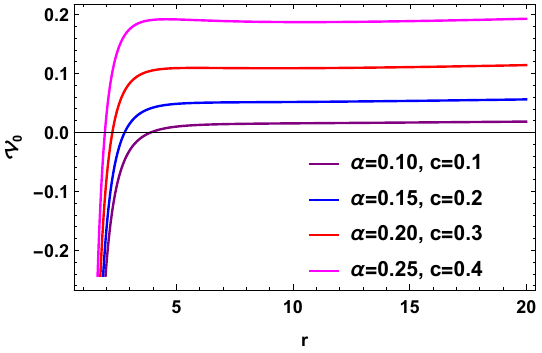}}\quad
    \subfloat[$m=1$]{\centering{}\includegraphics[width=0.48\linewidth]{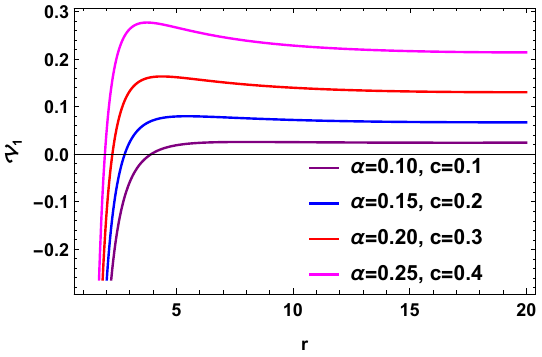}}
    \caption{Scalar perturbative potential $\mathcal{V}_m$ as a function of $r$.}
    \label{fig:scalar}
\end{figure}

\begin{figure}[ht!]
    \centering
    \subfloat[$m=0,\alpha=0.1$]{\centering{}\includegraphics[width=0.47\linewidth]{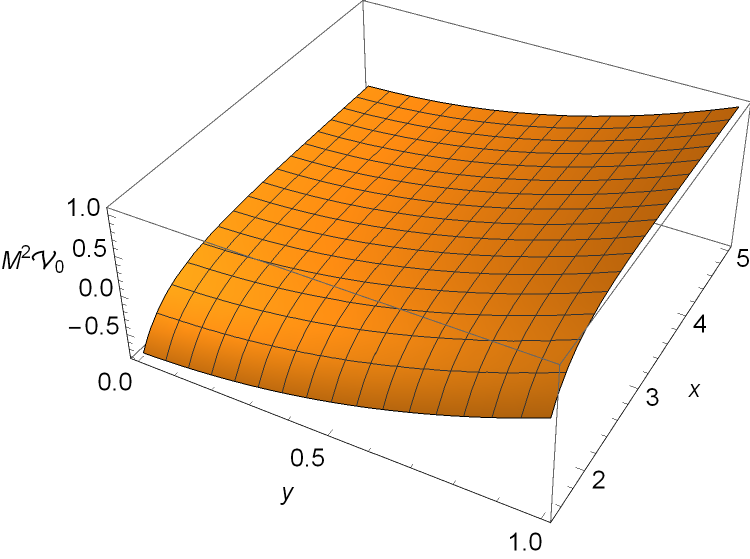}}\quad
    \subfloat[$m=0,\alpha=0.2$]{\centering{}\includegraphics[width=0.47\linewidth]{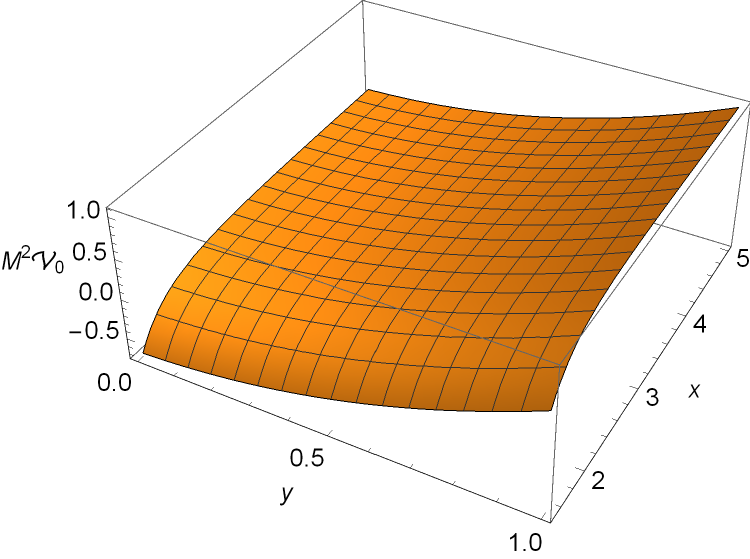}}\\
    \subfloat[$m=1,\alpha=0.1$]{\centering{}\includegraphics[width=0.47\linewidth]{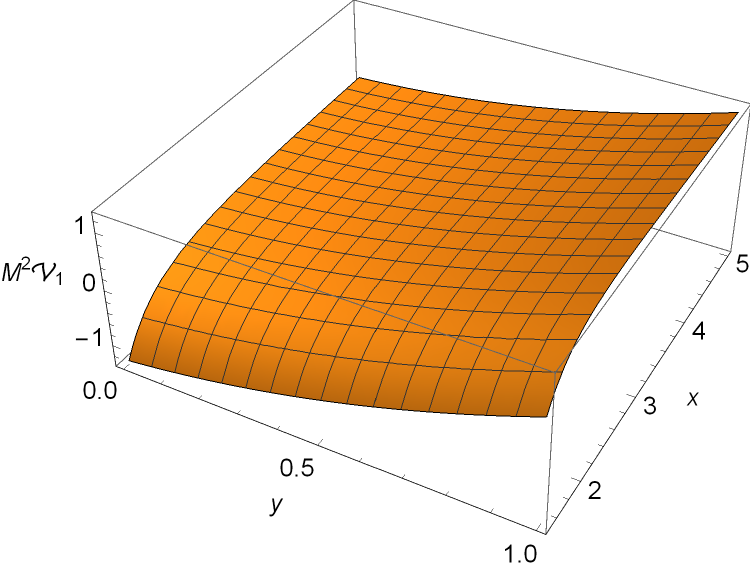}}\quad
    \subfloat[$m=1,\alpha=0.2$]{\centering{}\includegraphics[width=0.47\linewidth]{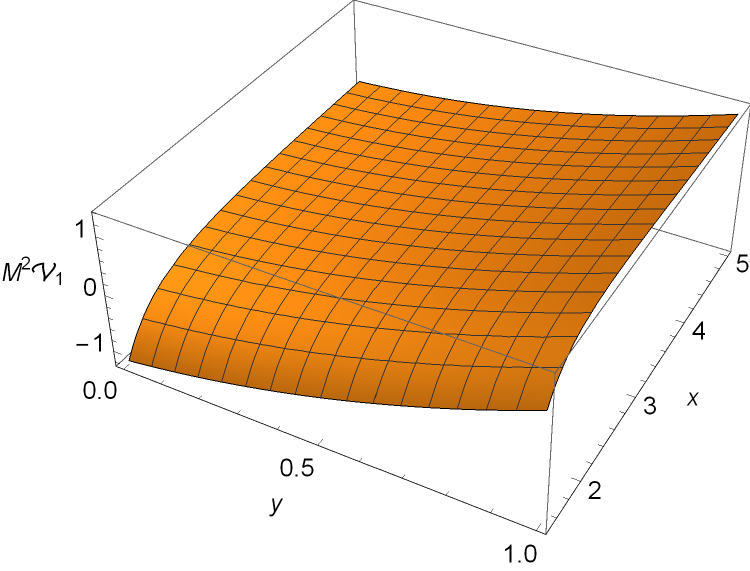}}
    \caption{There-dimensional illustration of the dimensional quantity $M^2\,\mathcal{V}_m$ for different values of $m$ and $\alpha$. Here, we set $M=1$.}
    \label{fig:three}
\end{figure}

In Figure \ref{fig:scalar}, we depict the scalar perturbative potential $\mathcal{V}_m$ given in Eq. (\ref{ff8}) as a function of $r$ varying the values of the CS parameter $\alpha$ and the QF constant $c$, while keeping the state parameter $w= -\frac{2}{3}$, the cosmological constant $\Lambda = -0.003$ and the arbitrary constant $k = 0.01$ fixed. {\color{black} We have observed that as the values of the CS parameter $\alpha$ and the quintessence constant $c$ both increased together, the scalar perturbative potential of the massless scalar field also increased with increasing $r$ for a particular value of $m=0$. A similar trend can also be seen for higher values of $m=1$. Increasing the scalar perturbation potential with distance $r$ means that the influence of the scalar field on the surrounding spacetime becomes stronger as we move farther away from the BH, at least within a certain range. It may also indicate that the scalar field is being ``trapped" or ``pinned" in certain regions of the spacetime, or the scalar field has long-range effects, influencing regions far from the BH. This phenomenon is important because it tells us how the scalar field interacts with the BH's gravitational field and how this interaction evolves as we move outward from the BH. In the context of BH thermodynamics, particularly Hawking radiation, a growing scalar perturbation potential could alter the way the BH evaporates. Scalar fields, for instance, could contribute to the energy emission from the BH or modify the spectrum of Hawking radiation.}

Similarly, in Figure \ref{fig:three}, we show the three-dimensional qualitative features of the dimensionless quantity $M^2\,\mathcal{V}_m$ given in Eq. (\ref{ff9}) for two values of $m=0\, \&\, 1$ and $\alpha=0.1\, \&\, 0.2$, while maintaining the state parameter $w = -\frac{2}{3}$, cosmological constant $\Lambda = -0.003$, arbitrary constant $k = 0.01$, and parameter $\mathrm{a} = M\,\sqrt{-\frac{\Lambda}{3}} = 0.0316$.

\subsection{\large \bf Thermodynamic Properties of A\lowercase{d}S BS Spacetime} \label{sec:6}

In this section, we study thermodynamics of the BS solution (\ref{aa1}) and analyze the effects of cosmic strings and QF on thermodynamic variables.

The BS mass can be obtained by solving $A(r=r_\text{h})=0$, in terms of the horizon radius $r_\text{h}$, as
\begin{equation}
    \alpha-\frac{2\,M_\text{h}}{r_\text{h}}+\frac{r^2_\text{h}}{\ell^2_{p}}+\frac{c}{r^{3\,w+1}_\text{h}}=0\Rightarrow M_\text{h}=\frac{1}{2}\,\left[\alpha\,r_\text{h}+\frac{r^3_\text{h}}{\ell^2_p}+\frac{c}{r^{3\,w}_\text{h}}\right].\label{hh1}
\end{equation}
Since the state parameter is in the range $-1 < w < -\frac{1}{3}$, thus, the solution to (\ref{hh1}) can vary according to it.  

Before proceedings for thermodynamic variables, let's us determine the event horizon radius $r_\text{h}$ for different state parameter.

\begin{center}
    {\bf Case A: State parameter, $w=-1/3$}
\end{center}

For the state parameter $w=-1/3$, the relation $A(r)=0$ implies
\begin{equation}
    \alpha-\frac{2\,M}{r}+\frac{r^2}{\ell^2_{p}}+c=0.\label{hh2}
\end{equation}
The single real root of the above equation is given by
\begin{equation}
    r_\text{h}=\Theta^{1/3}-\frac{\ell^2_{p}\,(\alpha+c)}{3}\,\Theta^{-1/3},\quad \Theta=\frac{1}{2}\,\left[\ell^2_{p}\,\sqrt{r^2_{s}+\frac{4}{27}\,(\alpha+c)^3\,\ell^2_{p}}+r^2_{s}\,\ell^2_{p} \right],\quad r_s=2\,M.\label{hh3}
\end{equation}
When the event horizon radius $r_\text{h}$ is considerably less than the AdS radius $|\ell_p|$, thus $r_{h} \ll |\ell_p|$, and neglecting the term $r^2_{h}/\ell^2_{p} \ll 1$ in Eq. (\ref{hh3}), the horizon radius becomes
\begin{equation}
    r_\text{h} = \frac{r_s}{\alpha+c} = \frac{2\,M}{\alpha+c}.\label{hh4}
\end{equation}

\begin{center}
    {\bf Case B: State parameter, $w=-2/3$}
\end{center}

For the state parameter $w=-2/3$, the relation $A(r)=0$ implies
\begin{equation}
    \alpha-\frac{2\,M}{r}+\frac{r^2}{\ell^2_{p}}+c\,r=0.\label{hh2aa}
\end{equation}
The solution to the above equation has a single real root, given by
\begin{eqnarray}
    &&r_\text{h}=\Pi^{1/3}+\left(\frac{c^2\,\ell^4_{p}}{9}-\frac{\alpha\,\ell^2_{p}}{3}\right)\,\Pi^{-1/3}-\frac{c\,\ell^2_{p}}{3},\nonumber\\
    &&\Pi=\frac{1}{2}\,\left[\sqrt{r^2_s+\frac{1}{27}\,(18\,\alpha\,c\,\ell^2_p-4\,c^3\,\ell^4_p)\,r_s-\frac{1}{27}\,\alpha^2\,c^2\,\ell^4_p+\frac{4}{27}\,\alpha^3\,\ell^2_p}+r^2\,\ell^2_p+\frac{\alpha\,c\,\ell^4_p}{3}-\frac{2}{27}\,c^3\,\ell^6_p \right].\label{hh5}
\end{eqnarray}

Employing the previous assumption that $r_h<<|\ell_p|$ and after simplification, we find
\begin{equation}
    r_\text{h}=\frac{\sqrt{4\,c\,r_s+\alpha^2}\pm \alpha}{2\,c}. \label{hh6}
\end{equation}
With the above result, one can write
\begin{equation}
    \frac{c\,r_\text{h}}{\alpha}=\frac{1}{2}\,\left(\sqrt{1+\frac{4\,c\,r_s}{\alpha^2}}-1\right).\label{hh7}
\end{equation}
Considering the term $\frac{c\,r_s}{\alpha^2}<<1$ and expanding up to the first, we find the horizon radius as follows
\begin{equation}
    r_\text{h} \approx \frac{r_s}{\alpha} \approx \frac{2\,M}{\alpha}.\label{hh8}
\end{equation}

\begin{center}
    {\bf Case C: State parameter, $w=-1$}
\end{center}

For the state parameter $w=-1$, the relation $A(r)=0$ implies
\begin{equation}
    \alpha-\frac{2\,M}{r}+\frac{r^2}{\ell^2_{p}}+c\,r^2=0.\label{hh9}
\end{equation}
The real root of the above equation is given by
\begin{eqnarray}
    &&r_\text{h}=\Delta^{1/3}-\frac{\alpha}{3\,\left(c+\frac{1}{\ell^2_p}\right)}\,\Delta^{-1/3},\nonumber\\
    &&\Delta=\left(\frac{\alpha}{3\,\left(c+\frac{1}{\ell^2_p}\right)}\right)^{3/2}\,\left[\sqrt{1+\frac{27}{4}\,\frac{\left(c+\frac{1}{\ell^2_p}\right)\,r^2_s}{\alpha^3}}+\sqrt{\frac{27}{4}\,\frac{\left(c+\frac{1}{\ell^2_p}\right)\,r^2_s}{\alpha^3}}\right].\label{hh10}
\end{eqnarray}

In the regime where $\frac{\left(c+\frac{1}{\ell^2_p}\right)\,r^2_s}{\alpha^3}<<1$, we can approximate $\Delta^{1/3}$ as
\begin{eqnarray}
    \Delta^{1/3}=\left(\frac{\alpha}{3\,\left(c+\frac{1}{\ell^2_p}\right)}\right)^{1/2}\,\left[1+\frac{27}{8}\,\frac{\left(c+\frac{1}{\ell^2_p}\right)\,r^2_s}{\alpha^3}+\frac{1}{3}\,\sqrt{\frac{27}{4}\,\frac{\left(c+\frac{1}{\ell^2_p}\right)\,r^2_s}{\alpha^3}}\right],\label{hh11}
\end{eqnarray}
and conversely
\begin{eqnarray}
    \Delta^{-1/3}&=&\left(\frac{\alpha}{3\,\left(c+\frac{1}{\ell^2_p}\right)}\right)^{-1/2}\,\left[1+\frac{27}{8}\,\frac{\left(c+\frac{1}{\ell^2_p}\right)\,r^2_s}{\alpha^3}-\frac{1}{3}\,\sqrt{\frac{27}{4}\,\frac{\left(c+\frac{1}{\ell^2_p}\right)\,r^2_s}{\alpha^3}}\right].\label{hh12}
\end{eqnarray}

Substituting the results given in Eqs. (\ref{hh11}) and (\ref{hh12}) into Eq. (\ref{hh10}), we obtain the horizon radius as follows
\begin{equation}\label{hh13}
    r_\text{h}= \frac{r_s}{\alpha} = \frac{2\,M}{\alpha}
\end{equation}

\begin{table}[ht!]
    \centering
    \begin{tabular}{|c|c|c|}
    \hline
    $w$ & $r_\text{h}$ & $M=\frac{1}{2}\,\left(\alpha\,r_\text{h}+\frac{r^3_\text{h}}{\ell^2_p}+\frac{c}{r^{3\,w}_\text{h}}\right)$\\ [1.5ex] 
    \hline
    -1 & $\frac{r_s}{\alpha}$ & $\frac{1}{2}\,\left(r_s+\frac{1}{\alpha^3}\,\left(\frac{1}{\ell^2_p}+c\right)\,r^3_s\right)$ \\[1.5ex] 
    \hline
    -2/3 & $\approx \frac{r_s}{\alpha}$ & $\approx \frac{1}{2}\,\left(r_s+\frac{r^3_s}{\alpha^3\,\ell^2_p}+\frac{c}{\alpha^2}\,r^2_s\right)$\\[1.5ex] 
    \hline
    -1/3 & $\frac{r_s}{\alpha+c}$ & $\frac{1}{2}\,\left(1+\frac{r^3_s}{(\alpha+c)^3\,\ell^2_p}\right)$\\[1.5ex] 
    \hline
    \end{tabular}
    \caption{Mass of BS for different state parameter}
    \label{tab:1}
\end{table}

The Hawking temperature is defined to be proportional to the surface gravity $\kappa$ by
 \begin{equation}
     T=\frac{\kappa}{2\,\pi},\quad\quad \kappa=A'(r_\text{h})/2.\label{jj4}
 \end{equation}
In our case, we find 
\begin{equation}
    T_H=\frac{1}{2\,\pi\,r_\text{h}}\,\left(\frac{M_\text{h}}{r_\text{h}}-\frac{c\,(3\,w+1)/2}{r^{3\,w+1}_\text{h}}+\frac{r^2_\text{h}}{\ell^2_p}\right).\label{jj5}
\end{equation}

\begin{figure}[ht!]
    \centering
    \subfloat[$w=-1$]{\centering{}\includegraphics[width=0.48\linewidth]{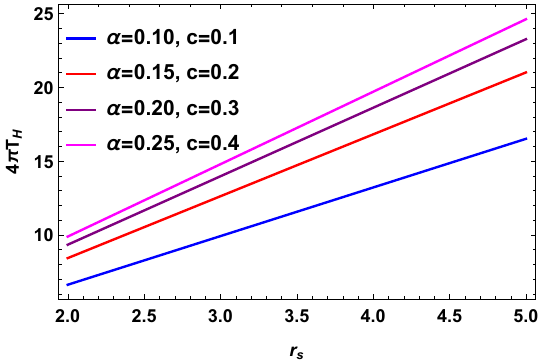}}\quad
    \subfloat[$w=-2/3$]{\centering{}\includegraphics[width=0.48\linewidth]{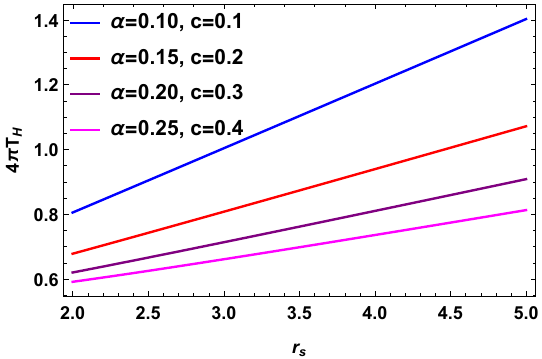}}
    \caption{The Hawking temperature $T_H$ as a function of the Schwarzschild-like radius $r=r_s$. Here $M=1$ and $\Lambda=-0.03$.}
    \label{fig:Hawking}
\end{figure}

\begin{table}[ht!]
    \centering
    \begin{tabular}{|c|c|c|c|}
    \hline
    $w$ & $r_\text{h}$ & $M_\text{h}=\frac{1}{2}\,\left(\alpha\,r_\text{h}+\frac{r^3_\text{h}}{\ell^2_p}+\frac{c}{r^{3\,w}_\text{h}}\right)$ & $T_H=\frac{1}{2\,\pi\,r_\text{h}}\,\left(\frac{M_\text{h}}{r_\text{h}}-\frac{c\,(3\,w+1)/2}{r^{3\,w+1}_\text{h}}+\frac{r^2_\text{h}}{\ell^2_p}\right)$\\ [1.5ex] 
    \hline
    -1 & $\frac{r_s}{\alpha}$ & $\frac{1}{2}\,\left(r_s+\frac{1}{\alpha^3}\,\left(\frac{1}{\ell^2_p}+c\right)\,r^3_s\right)$ & $\frac{\alpha^2}{4\,\pi\,r_s}\,\left[1+\frac{3}{\alpha^3}\,\left(c+\frac{1}{\ell^2_p}\right)\,r^2_s\right]$ \\[1.5ex] 
    \hline
    $-\frac{2}{3}$ & $\approx \frac{r_s}{\alpha}$ & $\approx \frac{1}{2}\,\left(r_s+\frac{r^3_s}{\alpha^3\,\ell^2_p}+\frac{c}{\alpha^2}\,r^2_s\right)$ & $\approx \frac{\alpha^2}{4\,\pi\,r_s}\,\left[1+\frac{2}{\alpha^3\,\ell^2_p}\,r^2_s+\frac{2\,c}{\alpha^2}\,r_s\right]$\\[1.5ex] 
    \hline
    $-\frac{1}{3}$ & $\frac{r_s}{\alpha+c}$ & $\frac{1}{2}\,\left(r_s+\frac{r^3_s}{(\alpha+c)^3\,\ell^2_p}\right)$ & $\frac{(\alpha+c)^2}{4\,\pi\,r_s}\,\left[1+\frac{3}{(\alpha+c)^3\,\ell^2_p}\,r^2_s\right]$\\[1.5ex] 
    \hline
    \end{tabular}
    \caption{Hawking Temperature at the event horizon radius. Here, $r_s=2\,M$.}
    \label{tab:2}
\end{table}

\begin{table}[ht!]
    \centering
    \begin{tabular}{|c|c|c|c|}
    \hline
    $w$ & $r_\text{h}$ & $\sigma$ & $S$\\[3ex] 
    \hline
    -1 & $\frac{r_s}{\alpha}$ & $\frac{2\,\pi\,r^2_s}{\alpha^2\,\ell^2_p}$ & $\frac{\pi\,r^2_s}{2\,\alpha^2\,\ell^2_p} $ \\[3ex] 
    \hline
    $-\frac{2}{3}$ & $\approx \frac{r_s}{\alpha}$ & $\approx \frac{2\,\pi\,r^2_s}{\alpha^2\,\ell^2_p}$ & $\approx \frac{\pi\,r^2_s}{2\,\alpha^2\,\ell^2_p}$\\[3ex] 
    \hline
    $-\frac{1}{3}$ & $\frac{r_s}{\alpha+c}$ & $\frac{2\,\pi\,r^2_s}{(\alpha+c)^2\,\ell^2_p}$ & $\frac{\pi\,r^2_s}{2\,(\alpha+c)^2\,\ell^2_p}$\\[3ex] 
    \hline
    \end{tabular}
    \caption{Area of the event horizon and Bekenstein-Hawking entropy for different state parameter. Here, $r_s=2\,M$.}
    \label{tab:3}
\end{table}
 
{\color{black} Figure \ref{fig:Hawking} illustrates the variation of the Hawking temperature for different values of the CS parameter $\alpha$ and the QF constant $c$, for two distinct state parameters, $w=-1$ and $w=-2/3$, respectively. It is observed that for a fixed value of the CS parameter $\alpha$ and the quintessence constant $c$, the Hawking temperature exhibits a linear increase as the Schwarzschild-like radius $r_s$ increases. In panel (a), this increasing trend of the Hawking temperature is enhanced when both the CS parameter $\alpha$ and the quintessence constant $c$ are increased simultaneously. In contrast, in panel (b), the increasing trend of the Hawking temperature shifts downward as the values of both the CS parameter $\alpha$ and the quintessence constant $c$ are raised. This behavior can be attributed to the rise in the QF state parameter $w$. One could also explore other values of the state parameter within the range $-1 < w < -2/3$ to further investigate how the Hawking temperature responds to these changes.}

The area of the event horizon of a BS is given by \cite{RGC}
\begin{equation}
\sigma=\frac{2\,\pi\,r^2_\text{h}}{\ell^2_p}.\label{jj6}    
\end{equation} 
Thus, the Bekenstein-Hawking entropy, in geometrized units, is given by
\begin{equation}
    S=\frac{\sigma}{4}=\frac{\pi\,r^2_\text{h}}{2\,\ell^2_p}.\label{jj7}
\end{equation}

\section{\large Conclusions}\label{sec:7}

In this study, we investigated the geodesic structure, shadow, perturbative dynamics, and thermodynamic properties of a cylindrically symmetric AdS BS surrounded by CS and QF. Our analysis provided a deeper understanding of the physical implications of these additional fields on the gravitational, perturbative, and thermodynamic behavior of the BS solution. We began by introducing the metric function for the AdS BS surrounded by CS and QF, given in Eq.~(\ref{aa1})--(\ref{aa2}), where the effects of the CS parameter $\alpha$ and the QF parameters $(c, w)$ were explicitly incorporated. The metric displayed deviations from the classical AdS BS solution due to these additional fields, significantly altering the causal structure of the spacetime.

Next, we examined the geodesic motion of both light-like and time-like test particles. The effective potential governing the trajectories of these particles was derived in Eq.~(\ref{bb5}), highlighting the modifications induced by the CS and QF parameters. The analysis of photon orbits allowed us to determine the CPO radius $r_{ph}$, expressed in Eq.~(\ref{cc4}) for $w=-2/3$ and Eq.~(\ref{cc6}) for $w=-1/3$, revealing that the CS and QF parameters shrink the CPO radius compared to the classical case. The influence of these parameters on the shadow radius of the BS was illustrated in Fig.~\ref{figa1}, which shows that increasing $\alpha$ and $c$ results in a reduction of the shadow size, an effect with potential astrophysical implications. The force acting on photon particles in the gravitational field of the BS was computed in Eq.~(\ref{cc6}), indicating that the CS and QF parameters directly affect the gravitational interaction experienced by massless particles. Moreover, the trajectory equation for photon motion was obtained in Eq.~(\ref{cc9}), further establishing how these external fields modify light propagation in this background. For time-like geodesics, we derived the conditions for stable circular orbits and obtained the expressions for specific angular momentum and energy of test particles, given in Eq.~(\ref{dd3}) and Eq.~(\ref{dd4}), respectively. The angular velocity of circular orbits was derived in Eq.~(\ref{dd5}), demonstrating its dependence on the CS and QF parameters.

The C-energy of the BS solution, a measure of gravitational energy contained within a cylindrical radius, was derived in Eq.~(\ref{ss9}). Our findings show that the C-energy density depends on both the CS and QF parameters, influencing the overall energy distribution of the spacetime. This result underscores the significance of external matter fields in shaping the gravitational characteristics of BSs. We further analyzed the propagation of a massless scalar field in this background by solving the Klein-Gordon equation. The perturbative potential $V(r)$, derived in Eq.~(\ref{ff7}), highlighted how the CS and QF parameters influence the stability of scalar field perturbations. The presence of these fields modified the effective potential, which directly affects quasinormal modes and stability properties of the BS spacetime.

The thermodynamic properties of the BS were extensively analyzed. We first determined the horizon radius $r_h$ for different values of $w$ by solving the relation $A(r)=0$. The mass function of the BS, given in Table~\ref{tab:1}, demonstrated how the CS and QF parameters modify the mass-radius relation. The Hawking temperature $T_H$, given in Eq.~(\ref{jj5}) and summarized in Table~\ref{tab:2}, showed that increasing $c$ and $w$ generally decreases the BS's temperature. The entropy, obtained via the Bekenstein-Hawking relation in Eq.~(\ref{jj7}), confirmed that the presence of CS and QF influences the thermodynamic behavior of the BS, as seen in Table~\ref{tab:3}. These results provide strong evidence that the presence of a CS and QF significantly alters the geodesic, perturbative, and thermodynamic properties of an AdS BS. The effects of $\alpha$ and $c$ manifest in deviations from standard results, leading to modifications in gravitational lensing, stability criteria, and thermodynamic phase structure.

{\color{black} Finally, it is worth noting that, while primarily theoretical, our work has several potential observational applications: (1) Our shadow radius calculations (Eqs. \ref{shadeq1}-\ref{bb16cc}) could inform future EHT observations of objects with potential cylindrical symmetry \cite{Cheng:2010ae,Johnson2023}; (2) The trajectory equation (\ref{solution}) provides distinctive gravitational lensing signatures that might be detected through precision astrometry \cite{Oikonomou2022}; (3) Our thermodynamic analysis, particularly the Hawking temperature's parameter dependence (Table \ref{tab:2}), connects to BH evaporation observations \cite{Barrau2019}; and (4) Our stability analysis of circular orbits could enhance understanding of accretion disk dynamics observable in X-ray data \cite{Cardoso2020, Chen2023}. Although speculative for cylindrically symmetric objects, our predictions could be tested as observational capabilities advance with next-generation instruments such as the Next Generation Event Horizon Telescope (ngEHT) and LISA \cite{Blackburn2019, Amaro-Seoane2023}.}

For future work, it would be valuable to extend this analysis to higher-dimensional BSs, incorporating additional interactions such as charge and rotation. A natural extension would involve investigating the quasinormal modes of the BS using numerical methods to assess the stability of the perturbative potential. Furthermore, a deeper analysis of phase transitions in this system, using methods such as holographic entanglement entropy, would provide insight into the thermodynamic nature of these modified BS solutions. Lastly, observational signatures of these modifications, particularly in the context of gravitational wave signals and lensing phenomena, could be explored in relation to upcoming astrophysical experiments \cite{Aurrekoetxea:2023vtp,Kumar:2023jgh,Vachher:2024fxs,Maji:2024cwv,AbhishekChowdhuri:2023ekr}.

{\small

\section*{Acknowledgments}

{\color{black} We would like to express our sincere gratitude to the editor and the anonymous referee for their valuable comments and constructive suggestions, which have significantly improved the quality and clarity of this manuscript.} F.A. acknowledges the Inter University Centre for Astronomy and Astrophysics (IUCAA), Pune, India for granting visiting associateship. \.{I}.~S. expresses gratitude to EMU, T\"{U}B\.{I}TAK, ANKOS, and SCOAP3 for their academic support. He also acknowledges COST Actions CA22113, CA21106, and CA23130 for their contributions to networking.

\section*{Data Availability Statement}
No data were generated or analyzed in this study. [Author’s comment: This is a theoretical work and no new data were generated and/or analyzed in the current study].

\section*{Code Availability Statement}
No code/software associated with this manuscript. [Author’s comment: No code/software were generated in the current study].
}

\end{document}